\documentclass[lettersize,journal]{IEEEtran}
\IEEEoverridecommandlockouts
\usepackage{tikz}
\usetikzlibrary{positioning, arrows.meta, calc, shapes.multipart, backgrounds}

\usepackage{cite}
\usepackage{amsthm}
\usepackage{amsmath,amssymb,amsfonts}
\usepackage{mathtools}
\usepackage{algorithm}
\usepackage{algpseudocode}
\usepackage{graphicx}
\usepackage{subcaption}
\usepackage{textcomp}
\usepackage[dvipsnames]{xcolor}

\usepackage{booktabs} 
\allowdisplaybreaks
\newtheorem{lemma}{Lemma}
\newtheorem{proposition}{Proposition}
\newtheorem{corollary}{Corollary}
\newtheorem{theorem}{Theorem}
\newenvironment{customlemma}[1]
  {\renewcommand\thelemma{#1}\lemma} 
  {\endlemma}
\newenvironment{customprop}[1]
  {\renewcommand\theproposition{#1}\proposition}
  {\endproposition}
\newenvironment{customcor}[1]
  {\renewcommand\thecorollary{#1}\corollary}
  {\endcorollary}
\newenvironment{customthm}[1]
  {\renewcommand\thetheorem{#1}\theorem}
  {\endtheorem}
\newcommand{\labelle}[1]{\overset{(#1)}{\le}}
\newcommand{\labeleq}[1]{\overset{(#1)}{=}}
\def\BibTeX{{\rm B\kern-.05em{\sc i\kern-.025em b}\kern-.08em
    T\kern-.1667em\lower.7ex\hbox{E}\kern-.125emX}}
\begin{document}

\title{
Data-Driven Deep MIMO Detection:\\ Network Architectures and Generalization Analysis
}

\author{\IEEEauthorblockN{Yongwei Yi,  Xinping Yi, \textit{Member, IEEE}, Wenjin Wang, \textit{Member, IEEE}, \\Xiao Li, \textit{Member, IEEE}, and Shi Jin, \textit{Fellow, IEEE}}
\thanks{The authors are with the School of Information Science and Engineering, 
Southeast University, Nanjing 210096, China. Emails:
\{yongweiyi, xyi, wangwj, li\_xiao, jinshi\}@seu.edu.cn.}
\thanks{Part of this work has been presented at the IEEE ICCC 2025~\cite{11148664}.}
}

\maketitle

\begin{abstract}
In practical Multiuser Multiple-Input Multiple-Output (MU-MIMO) systems, symbol detection remains challenging due to severe inter-user interference and sensitivity to Channel State Information (CSI) uncertainty. 
In contrast to the mostly studied belief propagation-type model-driven methods, which incur high computational complexity, Soft Interference Cancellation (SIC) strikes a good balance between performance and complexity.
To further address CSI mismatch and nonlinear effects, the recently proposed data-driven deep neural receivers, such as DeepSIC, leverage the advantages of deep neural networks for interference cancellation and symbol detection, demonstrating strong empirical performance. 
However, there is still a lack of theoretical underpinning for why and to what extent DeepSIC could generalize with the number of training samples.  

This paper proposes inspecting the fully data-driven DeepSIC detection within a Network-of-Multi-Layer Perceptrons (MLPs) architecture, which is composed of multiple interconnected MLPs via outer and inner Directed Acyclic Graphs (DAGs). Within such an architecture, DeepSIC can be upgraded as a graph-based message-passing process using Graph Neural Networks (GNNs), termed GNNSIC, with shared model parameters across users and iterations. Notably, GNNSIC achieves excellent expressivity comparable to DeepSIC with substantially fewer trainable parameters, resulting in improved sample efficiency and enhanced user generalization. By conducting a norm-based generalization analysis using Rademacher complexity, we reveal that an exponential dependence on the number of iterations for DeepSIC can be eliminated in GNNSIC due to parameter sharing.
Simulation results demonstrate that GNNSIC attains comparable or improved Symbol
Error Rate (SER) performance to DeepSIC with significantly fewer parameters and training
samples.
\end{abstract}

\section{Introduction}
\IEEEPARstart{M}{ultiuser} Multiple-Input Multiple-Output (MU-MIMO) has been a cornerstone technology for wireless communication systems, meeting the rapidly increasing throughput demands of modern wireless services. 
In uplink cellular networks, base stations equipped with multiple antennas can simultaneously serve multiple user terminals by exploiting spatial multiplexing, enabling significant improvements in spectral efficiency~\cite{foschini1998limits,marzetta2010noncooperative}.
Nevertheless, symbol detection in MU-MIMO systems is still a fundamental challenge, stemming from severe inter-user interference, which complicates the reliable recovery of transmitted symbols. 

Past decades have witnessed persistent efforts in addressing MU-MIMO detection, including the classical model-based and the recently emerged neural detection approaches.
Classical model-based detectors—such as maximum-likelihood (ML) detection and maximum a-posteriori (MAP) estimation—offer optimal symbol recovery but require exhaustive evaluation of all symbol combinations, leading to exponential complexity with respect to system dimension~\cite{paulraj2003introduction}. To reduce complexity, suboptimal alternatives such as symbol-wise detection simply treat interference as noise, yet inevitably sacrificing detection accuracy and throughput~\cite{el2011network}. These limitations have motivated the development of iterative interference cancellation schemes~\cite{andrews2005interference}, which refine soft symbol estimates across iterations using probabilistic reasoning. Such methods can closely approach MAP performance with only polynomial complexity by iteratively canceling estimated interference~\cite{che2015successive}, especially when enhanced with soft decoding~\cite{WJ-00, wang1999iterative, Alexander_1999}. However, their performance can degrade substantially in scenarios with imperfect Channel
State Information (CSI) or when the channel deviates from the linear Gaussian model.

To overcome these limitations, deep learning (DL)-based detectors have attracted
increasing attention for MU-MIMO symbol detection. In particular,
model-driven DL architectures based on deep unfolding have demonstrated
remarkable performance gains by embedding domain knowledge from classical
detectors into trainable neural networks
(see~\cite{shlezinger2023model,yu2025deep} and references therein). Representative
works include OAMP-Net2~\cite{he2020model}, which unfolds the orthogonal
approximate message passing (OAMP) detector~\cite{Ma2017Orthogonal} with a small
number of trainable parameters to enhance robustness under channel uncertainty; MMNet~\cite{9103314}, which is inspired by iterative soft-thresholding algorithms and
exploits temporal and spectral correlations in realistic channels to accelerate
training; 
GEPNet~\cite{9832663, zhou2023graph}, which incorporates Graph Neural Networks (GNNs) into the expectation propagation (EP) framework~\cite{Cespedes2014Expectation}
to improve message passing at the cost of increased computational complexity; and
AMP-GNN~\cite{10278093}, which further reduces complexity by unfolding the Approximate Message Passing (AMP) algorithm
\cite{Jeon2015Optimality} with embedded GNN modules, achieving a favorable
trade-off between complexity and Symbol Error Rate (SER). Despite their effectiveness, model-driven architectures remain fundamentally
constrained by the assumptions and limitations of the underlying algorithms, which makes the generalization with respect to different system and user configurations challenging.

These limitations have motivated the development of \emph{data-driven} neural
detectors. Representative works include the early DNN-based approach
DetNet~\cite{8642915}, which unfolds a projected gradient descent algorithm for
MIMO detection and demonstrates improved robustness compared to classical AMP
under certain channel conditions, but typically requires a large amount of
training data and suffers performance degradation at higher modulation orders;
ViterbiNet~\cite{shlezinger2020viterbinet}, which replaces the classical Viterbi
algorithm with a simplified DNN for detection over finite-memory channels, but
suffers from poor scalability with channel memory length; RE-MIMO~\cite{9298921}, which
further exploits recurrent and permutation-equivariant architectures to unfold
classical iterative detectors, achieving improved performance through attention-based message passing at the cost of increased model complexity and training
difficulty; ChannelNet~\cite{11113418}, which exploits
architectural symmetry through channel embedding and antenna-wise parameter
sharing to reduce complexity, yet typically requires a large number of
iterations to achieve high detection accuracy; and more recently SGT~\cite{hong2026softgraphtransformermimo}, which introduces
a graph-based transformer architecture to enhance message passing in iterative
detection, but entails substantial computational overhead and training cost due
to its large-scale attention modules. 

Of particular relevance to the current work is DeepSIC~\cite{shlezinger2020deepsic}, which
provides a representative neural realization of Soft Interference Cancellation
(SIC) by replacing both the interference cancellation and soft decoding
stages with trainable neural modules. DeepSIC infers soft symbol estimates from
the received signals without relying on explicit probabilistic modeling, while
preserving the iterative structure of classical SIC. Subsequent extensions of DeepSIC have investigated its robustness and
adaptability under more challenging channel conditions, including meta-learning
based schemes for handling rapidly varying channels~\cite{10041001} and
hypernetwork-based architectures for scalable and adaptive MIMO
detection~\cite{10830517}. As reported in prior work, DeepSIC and its extensions
achieve competitive detection performance across a wide range of MIMO
configurations, making them widely adopted baselines for learning-based
multiuser detection.

Despite excellent empirical performance, the success of DeepSIC naturally raises
questions regarding the model expressivity and generalization performance. For the former, is the size of neural networks sufficient to mimic the iterative SIC? For the latter, how many training samples are sufficient to train a model that is generalizable to different network and user configurations? 
With respect to the model expressivity, DeepSIC assigns independent neural modules to each user and each SIC iteration, which causes the number of trainable parameters to grow rapidly with the system dimension and
iteration depth. Although this design does not prevent strong empirical
performance in moderate-sized settings, it may limit sample efficiency and
generalizability when applied to larger or deeper SIC architectures.
With respect to the model generalization, the majority of existing works verify the models through extensive experiments, while there is still a lack of theoretical generalization analysis.

Generalization analysis aims to provide theoretical guarantees for learning algorithms with respect to data distributions, network architectures, model parameters, and training algorithms.
This line of research has recently attracted increasing attention in the wireless communication community.
Existing works have primarily focused on the learning of decoders or codebooks
under fixed channel models or additive noise assumptions,
establishing generalization bounds for decoder selection,
codebook design, and related surrogate losses
\cite{9618913,10041215}.
More recently, generalization analyses using Rademacher complexity have been developed
for neural belief propagation decoders operating on Tanner graphs,
explicitly characterizing the impact of code parameters,
graph connectivity, and iteration depth \cite{10418200}.
Of particular relevance is the generalization analysis in \cite{10121671} for deep learning-based MIMO detectors, showing that data-driven ReLU networks
can asymptotically approximate the MAP detector, with generalization error
controlled by the training sample size.

While these results provide important insights into the learnability of
decoder architectures in coded communication systems and deep MIMO detectors, they share a common underlying assumption that the computation graph
or functional structure is fixed \textit{a priori} or independent of the
physical channel realization.
In contrast, neural receivers for multi-user detection operate on
interference graphs induced by the physical channel itself, and rely on
user-level message passing for soft interference mitigation.
The resulting channel-induced interaction patterns, together with iterative
neural processing across multiple users and iterations, lead to a hierarchical computation structure that is not explicitly captured by existing theoretical
frameworks.
Although generalization analyses for deep MIMO detectors have been reported in \cite{10121671}, such results typically treat the neural receiver as a black-box function class
and provide limited insight into how architectural choices affect the
generalization behavior of practical neural multi-user receivers. These limitations call for a new analytical framework for generalization analysis that takes both network architecture and model parameters into account.

Motivated by these considerations, we propose to inspect the DeepSIC-type MU-MIMO detection from a new architectural perspective as a Network-of-Multi-Layer Perceptrons (MLPs), which is composed of a number of MLPs interconnected by outer and inner Directed Acyclic Graphs (DAGs). Within such a network architecture, DeepSIC can be seen as a special case, for which it can be further upgraded as a more compact network architecture with substantially reduced model parameters. The generalization analysis is also conducted for such a \emph{Network-of-MLPs}, revealing how network architectures and model parameters play a role in the required number of training samples for generalization.
The main contributions of this work are summarized as follows:
\begin{itemize}
    \item Within the architecture of \emph{Network-of-MLPs}, we propose \emph{GNNSIC}, a data-driven soft interference cancellation
    framework for MU-MIMO detection. The proposed method formulates the iterative SIC procedure as a graph-based message-passing process and employs
    GNN modules to perform interference cancellation and soft symbol detection.
    By imposing parameter sharing across users and SIC iterations, GNNSIC achieves a compact architecture with substantially fewer trainable parameters than DeepSIC. The extensive simulation results demonstrate that GNNSIC attains comparable or improved SER performance to DeepSIC in various linear/nonlinear channels with/without CSI uncertainty and possesses excellent user generalization performance, with significantly fewer parameters and training samples.
    \item We formulate a novel framework of norm-based generalization analysis for the \emph{Network-of-MLPs} architecture with Rademacher complexity, relating generalization performance to the network architectures and Frobenius norms of the model parameters.  
    The proposed \emph{Network-of-MLPs} consists of an outer DAG that captures the SIC iteration structure and inner DAGs that represent the MLP computations within each block. We develop a two-level peeling-based analysis that separately controls the inner and outer DAGs, enabling a compositional characterization of the overall
    Rademacher complexity.
    Such analyses have proven effective for deep architectures with compositional structures and parameter sharing, making them well-suited for studying learning-based iterative multi-user receivers with soft interference mitigation and shared neural components.
    \item Under the analytical generalization framework, we derive corresponding
    norm-based generalization bounds for both DeepSIC and GNNSIC. 
    For DeepSIC, the absence of parameter sharing across iterations leads to a
    multiplicative accumulation of block-wise operator  norms, while the unrolled outer DAG
    structure introduces degree-dependent amplification through repeated peeling,
    which together yield an exponential dependence of the complexity on the SIC depth.
    In contrast, for GNNSIC, extensive parameter sharing across both network blocks
    and outer iterations collapses the outer DAG to a single shared layer,
    thereby preventing both norm accumulation and degree amplification and
    yielding a substantially tighter bound independent of the SIC depth.
    This provides a theoretical explanation for the improved generalization behavior of
    GNNSIC observed in the numerical results.
\end{itemize}

The rest of this paper is organized as follows.
Section~\ref{sec:system_model-ISIC-DeepSIC} introduces the system model and reviews
iterative SIC and DeepSIC.
Section~\ref{sec:Network-of-MLPs} formalizes the \emph{Network-of-MLPs} abstraction and
shows how DeepSIC fits into this framework.
Section~\ref{sec:GNNSIC_framework} presents the proposed GNNSIC architecture.
Section~\ref{sec:generalization_analysis} develops the generalization analysis for
DeepSIC and GNNSIC.
Section~\ref{sec:numerical_results} reports the experimental results, and
Section~\ref{sec:conclusion} concludes the paper.

\section{System Model, Iterative SIC and DeepSIC}
\label{sec:system_model-ISIC-DeepSIC}
\subsection{System Model}
We consider an uplink multiuser communication system in which multiple single-antenna users simultaneously transmit their data over the same frequency resource to a multi-antenna base station. In this work, the communication system is modeled as symbol transmission over a linear channel characterized by the channel matrix $\boldsymbol{H}$, impaired by additive white Gaussian noise (AWGN), i.e.,
\begin{align}
    \boldsymbol{y} = \boldsymbol{H}\boldsymbol{s}+\boldsymbol{w},
    \label{eq:channel_out}
\end{align}
where $\boldsymbol{y} \in \mathbb{R}^{N}$ is the channel output, representing the received signal at the base station. $\boldsymbol{H} = [\boldsymbol{h}_1, \boldsymbol{h}_2, \dots, \boldsymbol{h}_K] \in\mathbb{R}^{N\times K}$ is the channel matrix, where $\boldsymbol{h}_i$ is the user-$i$'s channel, encapsulating the channel coefficients between the $i$-th user and the $N$ receiver antennas at the base station. The users' channels are assumed to be constant during communication and known to the receiver. $\boldsymbol{s}=(s_1, s_2, \dots, s_K) \in \mathcal{M}^{K}$ represents the input symbols transmitted by the $K$ users. Each element of $\boldsymbol{s}$ is drawn from a constellation set $\mathcal{M}$ with size $M$, where $M$ denotes the number of constellation points (e.g., for BPSK, $M = 2$), and $\boldsymbol{w} \in \mathbb{R}^{N}$ is an independent multivariate Gaussian noise vector with zero mean and covariance $ \sigma_n^2 \mathbf{I}_N$.

Different from the most popular line of research that is dedicated to model-based/driven MIMO detection, we place our focus on the iterative SIC and its data-driven counterpart.

\subsection{Iterative SIC}
The iterative SIC algorithm \cite{WJ-00} is to enhance multiuser detection by iteratively canceling interference across multiple stages and making soft decisions. In each iteration, the conditional distribution of the input symbol for a given user, $s_i$, is estimated based on the channel output $\boldsymbol{y}$, by utilizing previous estimates of the conditional distributions of the interfering symbols. By recursively applying this process, the estimates of the conditional distributions are progressively refined, so that the final detection of the input symbols is obtained based on the conditional distributions in the final iteration.

The iterative SIC algorithm involves $L$ iterations, each consisting of two main steps performed in parallel: interference cancellation and soft detection.

\begin{enumerate}
    \item \textbf{Interference Cancellation}: For the $i$-th user and the $l$-th iteration, interference cancellation starts by estimating the interference from other users $k \neq i$. The interference generated by other users can be collected as $\{\boldsymbol{h}_k s_k\}_{k \neq i}$.
    Since $s_k$ cannot be directly observed by the receiver, an estimate at the previous iteration, $\hat{s}_k^{(l-1)}$, can be obtained through its expected value $\mathbb{E}[\hat{s}_k^{(l-1)}]$, where $\hat{s}_k^{(l-1)}$ takes all possible constellation points in $\mathcal{S}$ with the probabilities provided by the previous iteration $\hat{\boldsymbol{p}}_k^{(l-1)}$, and the variance $\mathrm{Var}(\hat{s}_k^{(l-1)})$. Therefore, the interference from other users can be subtracted from the received channel output $\boldsymbol{y}$, where the symbols $\{s_k\}_{k \neq i}$ are replaced by their expected estimate, i.e., $\{\mathbb{E}[\hat{s}_k^{(l-1)}]\}_{k \neq i}$.

    \item \textbf{Soft Detection}: In the soft detection step, the input symbols for user $i$ at iteration $l$ are estimated based on the conditional distribution $\hat{\boldsymbol{p}}_i^{(l)}$. Let $\boldsymbol{z}_i^{(l)} \in \mathbb{R}^N$ represent the updated channel output
after interference cancellation, which is given by
\begin{align}
\boldsymbol{z}_i^{(l)}
=
\boldsymbol{h}_i s_i
+
\sum_{k \neq i}
\boldsymbol{h}_k \Delta s_k^{(l-1)}
+
\boldsymbol{w},
\end{align}
where
$\Delta s_k^{(l-1)}$ denotes the residual estimation error of the interfering
symbol $s_k$ at iteration $(l-1)$, defined as
\(
\Delta s_k^{(l-1)}
=
s_k - \mathbb{E}[\hat{s}_k^{(l-1)}].
\) Thus, the conditional distribution of $\boldsymbol{z}_i^{(l)}$, given the input symbol $s_i$ set to a constellation point $\alpha_{m}\in\mathcal{M}$, corresponds to a multivariate Gaussian distribution with mean value $\boldsymbol{h}_i\alpha_{m}$ and covariance $\sum_{k \ne i}\boldsymbol{h}_k\boldsymbol{h}_k^T\mathrm{Var}(\hat{s}_k^{(l-1)}) + \sigma_n^2\mathbf{I}_N$. For each constellation point $\alpha_{m}$, the estimated conditional distribution $\hat{\boldsymbol{p}}_i^{(l)}$ is approximated using Bayes' theorem. After the final iteration, the detected symbol is the one that maximizes the conditional distribution $\hat{\boldsymbol{p}}_i^{(L)}$ for each user, i.e., $\hat{s}_i = \alpha_{m^*}$ where
    $
    m^* =
            \underset{m \in \{1, \ldots, M\}}{\arg\max}\left(\hat{\boldsymbol{p}}_{i,m}^{(L)}\right).
        \label{eq:hard_decision}
    $
\end{enumerate}
\subsection{DeepSIC}
DeepSIC \cite{shlezinger2020deepsic} is a data-driven implementation based on the iterative SIC algorithm. This method utilizes a series of interconnected neural network blocks to perform both the interference cancellation and soft detection stages inherent in the iterative SIC process. By transforming these sequential stages into a classification task, DeepSIC leverages the power of neural networks to enhance performance and robustness in complex signal environments.

Each network block in DeepSIC corresponds to a specific user and iteration. For the $ i $-th user at the $ l $-th iteration, the estimated conditional distribution $ \hat{\boldsymbol{p}}_i^{(l)} $ for symbol $s_i $ is computed based on the realization of the channel output $ \boldsymbol{y} $ and the conditional distribution estimates obtained from the previous iteration for other users. This classification task is implemented using fully connected neural networks with a softmax layer, where the input features are the channel output, and the training label is the input symbol $s_i $ for user $ i $.

DeepSIC's architecture consists of multiple network blocks working in tandem. Specifically, each building block performs both interference cancellation and soft decoding for a given user. The output of each block is the estimated conditional probabilities of input symbols, denoted as $ \hat{\boldsymbol{p}}_i^{(l)} $ for user $ i $ at iteration $ l $. For a system with $ K $ users and $ L $ iterations, DeepSIC is constructed with $ K \times L $ network blocks, providing comprehensive and iterative predictions of the input symbols.
\indent The neural network blocks are trained using supervised learning. The training data consists of pairs of input features and corresponding transmitted symbols. The network is trained to minimize the cross-entropy loss between the predicted and actual symbol distributions. This training process enables the network to learn to accurately predict the conditional probabilities of the transmitted symbols.

Compared to the iterative SIC, DeepSIC promises excellent detection performance for both linear and non-linear channels as well as enhanced robustness to CSI uncertainty. However, DeepSIC requires a large number of received signal samples to train and retrain the models when the number of users varies. This is the motivation of our current work.
\begin{figure*}[htbp]
  \centering
  \includegraphics[width=0.9\textwidth]{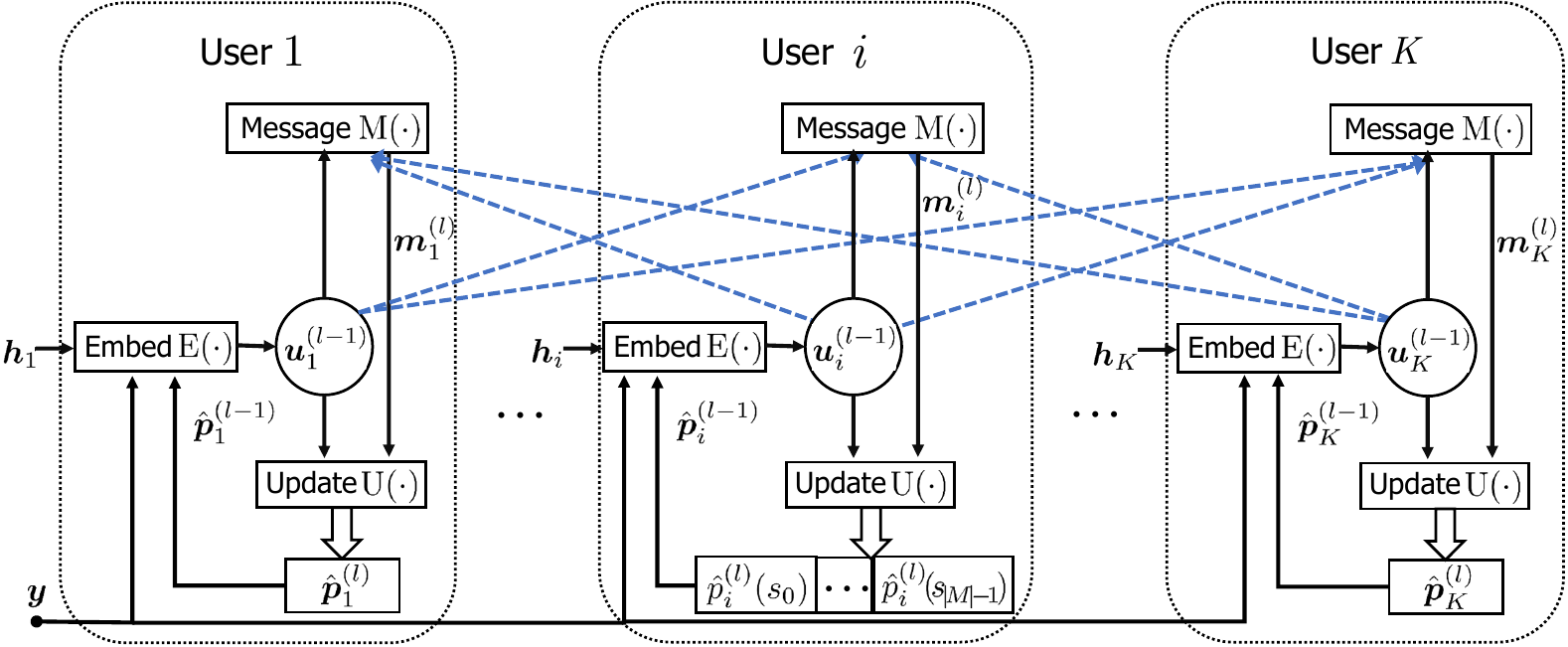}
  \caption{The internal structure of a GNNSIC network block at the $l$-th iteration.
Each user node forms its block input by concatenating the received signal
$\boldsymbol{y}$, the channel state $\boldsymbol{h}_i$, and the prior soft symbol estimate
$\hat{\boldsymbol{p}}_i^{(l-1)}$, which are processed by a set of MLPs constituting the
inner computation graph. Blue dashed lines indicate message exchanges among user
nodes induced by the underlying interference graph. This block is repeatedly 
applied across iterations according to the outer DAG defined by the 
\emph{Network-of-MLPs} architecture.}
  \label{fig:overall}
\end{figure*}

\section{Network-of-MLPs}
\label{sec:Network-of-MLPs}
In what follows, we reformulate DeepSIC in a general \emph{Network-of-MLPs} manner, which facilitates a new network architecture design in Section \ref{sec:GNNSIC_framework} and generalization analysis in Section \ref{sec:generalization_analysis}. 
\subsection{Network-of-MLPs: Architecture and Notation}
\label{sec:notation_network_of_mlp_arch}
\paragraph*{Notations}
Throughout the paper, expressions of the form
$(x_\mu)_{\mu=1}^b$ denote an ordered collection (i.e., a tuple or sequence)
indexed by $\mu$, rather than an unordered set.

We introduce a unified architecture, referred to as a
\emph{Network-of-MLPs}, which serves as a common representation for a broad class
of iterative neural detectors. The outer architecture is represented by a
directed acyclic graph (DAG) $G^{(\mathrm{out})}=(\mathcal{V}^{(\mathrm{out})},\mathcal{E}^{(\mathrm{out})})$, whose nodes correspond
to multiple \emph{network blocks}. We index the network blocks by user
$k\in[K]\triangleq\{1,\dots,K\}$ and outer iterations by $l\in[L]$.
Each network block is itself a Multilayer Perceptron (MLP) and therefore
induces an inner DAG. Each inner MLP block $\mathcal{B}_{k_l}$ is associated with a
directed acyclic graph $G^{(\mathrm{in})}_{k_l}=(\mathcal{V}^{(\mathrm{in})}_{k_l},\mathcal{E}^{(\mathrm{in})}_{k_l})$ composed of
$N+1$ layers indexed by $n=0,\dots,N$, where $n=0$ corresponds to the input layer.
From this perspective, the overall model can be viewed as a
\emph{DAG-of-DAGs}, where each MLP block is a super-node in the
outer graph.

The architecture is organized hierarchically, with
an outer structure indexed by network block $k$ and iteration $l$,
and an inner MLP structure of depth $N$ within each block. Each block processes local input features and
produces an intermediate soft estimate. The $j$-th neuron in layer $n$ is denoted by $\boldsymbol{z}^{(\mathrm{in}),n}_{k_l,j}$,
and we write $\boldsymbol{v}^{(\mathrm{in}),n}_{k_l}=(\boldsymbol{z}^{(\mathrm{in}),n}_{k_l,j})_{j=1}^{q^{(\mathrm{in})}_{k_l,n}}$ 
for the collection of neurons in that layer. The width of layer $n$ is $q^{(\mathrm{in})}_{k_l,n}$, and $c_n$ denotes the number of channels at that layer. Computation in $\mathcal{B}_{k_l}$ is parameterized by layer-wise weight matrices $\boldsymbol{W}^{(\mathrm{in}),1}_{k_l},\dots,\boldsymbol{W}^{(\mathrm{in}),N}_{k_l}$. For $1\le n\le N-1$, $\boldsymbol{W}^{(\mathrm{in}),n}_{k_l,j}$ is the weight matrix associated with the $j$-th neuron in layer $n$ and maps the concatenated activations of its predecessor neurons to $\mathbb{R}^{c_n}$. The final matrix $\boldsymbol{W}^{(\mathrm{in}),N}_{k_l}\in\mathbb{R}^{M\times q^{(\mathrm{in})}_{k_l, N-1}}$ implements a linear readout producing an $M$-dimensional output. For any neuron $\boldsymbol{z}^{(\mathrm{in}),n}_{k_l,j}$, $\mathrm{pred}(n,j)_{k_l}$ denotes the set of neurons in layer $n-1$ connected to it by directed edges in $\mathcal{E}_{k_l}$. For a fixed outer iteration $l$, the layer-wise sparsity of the inner DAG
for block $\mathcal{B}_{k_l}$ is defined as
\(
d^{(\mathrm{in})}_{k_l,n}
\triangleq
\max_{j\in[q^{(\mathrm{in})}_{k_l, n}]}
|\mathrm{pred}(n,j)_{k_l}|,
\)
and the maximum inner sparsity at outer iteration $l$ is given by
\(
d^{(\mathrm{in})}_{l,n}
\triangleq
\max_{k\in[K]}
d^{(\mathrm{in})}_{k_l,n}.
\) Dependencies across blocks are realized
through structured interconnections—such as feature concatenation or aggregation—which encode the algorithmic interactions across network blocks and iterations. At outer layer $l$, the collection of network blocks is denoted by
$\boldsymbol{v}^{(\mathrm{out}),l}=(\boldsymbol{z}^{(\mathrm{out}),l}_{k_l})_{k_l=1}^{q^{(\mathrm{out})}_l}$,
where $q^{(\mathrm{out})}_l$ is the width of that layer and
$\boldsymbol{z}^{(\mathrm{out}),l}_{k_l}$ denotes the $k_l$-th block.
Each block produces a vector-valued representation in $\mathbb{R}^{c_l}$,
where $c_l$ denotes the number of channels at outer iteration $l$.
The transformations across outer layers are parameterized by weight matrices
$\boldsymbol{W}^{(\mathrm{out}),1},\dots,\boldsymbol{W}^{(\mathrm{out}),L}$, where
$\boldsymbol{W}^{(\mathrm{out}),l}_{k_l}\in\mathbb{R}^{c_l\times (c_{l-1}\cdot|\mathrm{pred}(l,k_l)|)}$
aggregates the representations from the predecessor blocks in layer $l-1$
according to the connectivity pattern defined by $\mathcal{E}^{(\mathrm{out})}$.
For a block $\boldsymbol{z}^{(\mathrm{out}),l}_{k_l}$, we denote by $\mathrm{pred}(l,k_l)$
the set of its predecessor blocks in the $(l-1)$-th outer layer.
The layer-wise sparsity of the outer DAG is quantified by the maximum in-degree 
\(
     d^{(\mathrm{out})}_{l}
    \triangleq
    \max_{k_l\in[q^{(\mathrm{out})}_l]}
    \bigl|\mathrm{pred}(l,k_l)\bigr|.
\)
\paragraph*{Global degree parameters}
To characterize the overall sparsity of the \emph{Network-of-MLPs}, we define
\[
\bar{d}^{(\mathrm{out})}
\triangleq
\max_{l\in[L]}d^{(\mathrm{out})}_l,
\qquad
\bar{d}^{(\mathrm{in})}
\triangleq
\max_{l\in[L],\,n\in[N-1]}d^{(\mathrm{in})}_{l,n}.
\]
These quantities serve as global descriptors of the outer- and inner-level 
DAG sparsity and will be used to express the final complexity bounds.
\paragraph*{Input energy assumption}
Assume there exists a constant $E_0 > 0$ such that
\begin{align*}
\max_{k_{0}}
\sum_{t=1}^{T}
\big\|
\boldsymbol{z}^{(\mathrm{out}),0}_{k_{0}}(\boldsymbol{x}_t)
\big\|_2^2 \le E_0 .
\end{align*}
Formally, the architecture comprises $K$ network blocks unrolled over $L$ outer
iterations, resulting in a total of $K \times L$ MLP blocks and inducing a
DAG-of-DAGs computation graph. This abstraction separates the local nonlinear
transformations within individual MLPs from the global information propagation
across network blocks and iterations. Both DeepSIC and GNNSIC can be interpreted
as specific instantiations of this framework through different choices of block
connectivity and parameter sharing, which enables a unified analysis of model
complexity and generalization behavior in subsequent sections.

\subsection{DeepSIC as Network-of-MLPs}
DeepSIC~\cite{shlezinger2020deepsic} can be naturally interpreted as a specific
instantiation of the proposed \emph{Network-of-MLPs} architecture, obtained by
unrolling the classical iterative SIC
algorithm and replacing its analytical processing steps with learned neural
modules.

In DeepSIC, each \emph{network block} corresponds to a particular user and a
particular SIC iteration. For the $i$-th user at iteration $l$, the associated
MLP block produces an estimate of the conditional symbol distribution
$\hat{\boldsymbol{p}}_i^{(l)}$ based on the received signal $\boldsymbol{y}$ and
the soft symbol estimates of the remaining users from the previous iteration.
Within the \emph{Network-of-MLPs} abstraction, this corresponds to an outer DAG in which
blocks at iteration $l$ receive inputs from multiple blocks at iteration $l-1$
through feature concatenation.

Each block is implemented as a fully connected MLP with a softmax output layer,
thereby realizing a local classification task. The input features consist of the
channel observation together with the concatenated soft estimates from other
users, while the training label is the transmitted symbol $s_i$ of the
corresponding user. For a system with $K$ users and $L$ SIC iterations, the
resulting architecture contains $K \times L$ MLP blocks, with no parameter
sharing across users or iterations.

From a structural perspective, DeepSIC therefore corresponds to a
\emph{Network-of-MLPs} with dense inter-block connectivity induced by feature
concatenation and with independently parameterized blocks. While this design
offers strong empirical performance for both linear and nonlinear channels, it
also leads to a hypothesis space whose complexity grows rapidly with the number
of users and iterations. As will be analyzed in Section~\ref{sec:generalization_analysis}, 
this lack of parameter sharing plays a central role in the generalization
behavior of DeepSIC and motivates the structured and shared architecture adopted
by GNNSIC.

\section{GNNSIC Framework}
\label{sec:GNNSIC_framework}

This section presents GNNSIC, a graph-based neural MIMO detection framework that
can be interpreted as a structured instantiation of the
\emph{Network-of-MLPs} architecture introduced earlier. Similar to DeepSIC,
GNNSIC follows the iterative SIC paradigm while employing learnable nonlinear
processing. Unlike DeepSIC, GNNSIC explicitly models inter-user interference
through a conflict graph, which induces a structured outer DAG over a collection
of neural network blocks. This topology-aware design enables efficient
representation of inter-user dependencies and facilitates extensive parameter
sharing across users and iterations, resulting in a substantially reduced model
complexity.

\subsection{Overview of GNNSIC}

From the \emph{Network-of-MLPs} perspective, GNNSIC can be viewed as a collection of
network blocks organized over an outer DAG indexed by users and iterations.
Each block implements a fixed inner computation graph, realized by a set of
MLPs that perform local nonlinear processing and themselves form a DAG. The outer DAG then specifies how the outputs of these inner
graphs are propagated across blocks through explicit feature concatenation.

\subsubsection{Topology-aware Modeling of Multiuser Interference}

GNNSIC represents multiuser interference using a fully connected, bidirectional
graph \( \mathcal{G}^{(\mathrm{phy})} = (\mathcal{V}^{(\mathrm{phy})}, \mathcal{E}^{(\mathrm{phy})}) \), where each node
\( v^{(\mathrm{phy})}_i \in \mathcal{V}^{(\mathrm{phy})} \) corresponds to a user, and each directed edge
\( (v^{(\mathrm{phy})}_i, v^{(\mathrm{phy})}_j) \in \mathcal{E}^{(\mathrm{phy})} \) encodes the interference from user \( i \) to user
\( j \). Within the \emph{Network-of-MLPs} abstraction, each GNN iteration corresponds
to one application of a shared network block, while the graph structure specifies
the outer-DAG connectivity among blocks.

A shared GNN module is iteratively applied over \( L \) iterations to refine the
symbol estimates. The
structure of the \( l \)-th iteration is depicted in Fig.~\ref{fig:overall}. All iterations share an identical set of parameters, such that
the iterative procedure can be interpreted as an unrolling of the same network
block over the outer DAG, rather than a stack of independent layers. Accordingly,
each iteration corresponds to one unfolding of the outer DAG in terms of
information flow, while not increasing the model capacity. Information is
propagated across iterations by directly concatenating the current soft symbol
estimates with the received signal and channel features. The complete GNNSIC procedure is summarized
in Algorithm~\ref{alg:GNNSIC}.
\subsubsection{Parameter Sharing Strategies}

Within the \emph{Network-of-MLPs} framework, the key architectural novelty of GNNSIC
lies in its extensive parameter sharing across both the iteration and user
dimensions, which significantly enhances model compactness and generalizability.

\paragraph*{Cross-Iteration Sharing}
All network blocks across iterations share an identical set of parameters.
Specifically, the MLPs used for node embedding, message generation, and symbol
update are reused at every iteration. This recursive application of a single
shared block corresponds to repeatedly traversing the outer DAG, with
intermediate representations from the previous iteration concatenated into the
input of the next block. Such a design decouples the number of inference iterations from the model
capacity: although the block is unrolled for \( L \) iterations during inference,
the effective depth of the hypothesis class is determined by a single shared
network block.

\paragraph*{Cross-User Sharing}
Within each iteration, all user nodes employ the same embedding, message, and
update functions. From the \emph{Network-of-MLPs} viewpoint, this implies that all
network blocks associated with different users are parameter-tied, differing
only in their local input features. This user-agnostic design enables GNNSIC to
generalize across varying numbers of users without retraining, while preserving
a consistent outer-DAG structure. Importantly, such parameter tying prevents the
effective model capacity from scaling with the number of users, a property that
will be formally reflected in the generalization analysis.

Through both cross-iteration and cross-user parameter sharing,
GNNSIC realizes a highly compact \emph{Network-of-MLPs} architecture, achieving reduced
model complexity and memory usage while retaining strong expressive power.
\begin{algorithm}[t]
\caption{GNNSIC}
\label{alg:GNNSIC}

\textbf{Input:} Channel output $\boldsymbol{y}$; channel matrix $\boldsymbol{H}$; noise variance $\sigma_w^2$. \\[1mm]
\textbf{Initialization:}
\begin{itemize}
    \item Define the graph $\mathcal{G}^{(\mathrm{phy})}=(\mathcal{V}^{(\mathrm{phy})}, \mathcal{E}^{(\mathrm{phy})})$, where $\mathcal{V}^{(\mathrm{phy})}$ denotes the set of users (nodes) and $\mathcal{E}^{(\mathrm{phy})}$ the set of edges.
    \item Initialize the symbol posterior of each node $i \in \mathcal{V}^{(\mathrm{phy})}$ as
    \[
        \hat{\boldsymbol{p}}_i^{(0)} = \frac{1}{M} \mathbf{1} \in \mathbb{R}^M.
    \]
\end{itemize}

\textbf{Main Procedure:}
\begin{algorithmic}[1]
\For{$l = 1,\ldots,L$}
    \State $\boldsymbol{u}_i^{(l-1)} \gets \mathrm{E}\!\left(\boldsymbol{y}, \boldsymbol{h}_i, \hat{\boldsymbol{p}}_i^{(l-1)}\right)$
    \For{$i \in \mathcal{V}$}
        \State 
        $\boldsymbol{m}_i^{(l)} \gets 
        \mathrm{mean}\Big(
            \big\{
                \mathrm{M}\big(
                    \boldsymbol{u}_i^{(l-1)}, 
                    \boldsymbol{u}_k^{(l-1)}, 
                    \boldsymbol{e}_{ik}^{(l-1)}
                \big)
                \,\big|\,
                k \in \mathcal{N}(i)
            \big\}
        \Big)$
        \State $\hat{\boldsymbol{p}}_i^{(l)} \gets 
        \mathrm{U}\!\left(\boldsymbol{u}_i^{(l-1)}, \boldsymbol{m}_i^{(l)}\right)$
    \EndFor
\EndFor
\end{algorithmic}

\textbf{Output:} Final symbol estimate via hard decision:
\begin{equation}
    m_i^* = \arg\max_{m \in \{1,\ldots,M\}} 
    \hat{\boldsymbol{p}}_{i,m}^{(L)}, 
    \quad \forall i \in \mathcal{V}.
    \tag{7}
\label{eq:hard_decision1}
\end{equation}
\end{algorithm}

\subsection{GNNSIC for MIMO Detection}

\subsubsection{Initialization of Symbol Estimates}

At the beginning of the detection process, the initial soft estimate for each user \( i \) is set to a uniform distribution over the modulation alphabet, i.e., \( \hat{\boldsymbol{p}}_i^{(0)} = \frac{1}{M} \cdot \boldsymbol{1} \in \mathbb{R}^M \), where \( M \) denotes the modulation order and \( \boldsymbol{1} \) is an all-one vector, assuming equal probability for all symbols in the constellation \( \mathcal{M}^M \).

\subsubsection{GNN Module at Iteration \(l\)}

Each GNN module, corresponding to a single network block in the proposed
\emph{Network-of-MLPs} framework, consists of three stages: {node feature embedding}, {message passing}, and {symbol estimate updating}.

\paragraph*{a) Node Feature Embedding}

At the beginning of the \( l \)-th iteration, each user node \( i \) constructs its input feature vector by concatenating the received signal \( \boldsymbol{y} \), the channel vector \( \boldsymbol{h}_i \), and the soft symbol estimate from the previous iteration \( \hat{\boldsymbol{p}}_i^{(l - 1)} \). This concatenated vector is then transformed into a higher-dimensional latent representation via a shared embedding function:
\begin{align}
\boldsymbol{u}_i^{(l-1)} = \mathrm{E}\left(\boldsymbol{y}, \boldsymbol{h}_i, \hat{\boldsymbol{p}}_i^{(l - 1)}\right),
\end{align}
where \( \mathrm{E}: \mathbb{R}^{2N+M} \to  \mathbb{R}^{2a}\) denotes an MLP shared across users and iterations
with \( a \) being the latent dimensionality.

\paragraph*{b) Message Passing}

Each node aggregates messages from its neighbors as follows:
\begin{align}
\boldsymbol{m}_{i}^{(l)} 
= \mathrm{mean}\left( 
    \left\{ 
        \mathrm{M}\left(
            \boldsymbol{u}_{i}^{(l-1)}, 
            \boldsymbol{u}_k^{(l-1)}, 
            \boldsymbol{e}_{ik}^{(l-1)}
        \right) 
        \,\middle|\, 
        k \in \mathcal{N}(i) 
    \right\} 
\right),
\end{align}
where \( \mathrm{M}(\cdot) \) is a message generation function with a shared MLP across users and iterations followed by a softmax activation, and \( \boldsymbol{e}_{ik}^{(l-1)} \) denotes the edge feature between users \( i \) and \( k \). 
The generated messages from user $i$'s neighbors, i.e., for all $k \in \mathcal{N}(i)$, are aggregated at user $i$ with a $\mathrm{mean}(\cdot)$ function, and then passed back to each neighbor $k \in \mathcal{N}(i)$.

\paragraph*{c) Symbol Estimate Updating}

Each user $i$ then updates its symbol-wise posterior estimate according to the passed messages and its node feature embedding, i.e.,
\begin{align}
\hat{\boldsymbol{p}}_i^{(l)} = \mathrm{U}\left(\boldsymbol{u}_i^{(l-1)}, \boldsymbol{m}_i^{(l)}\right),
\end{align}
where \( \mathrm{U}: \mathbb{R}^{2a + M} \to \mathbb{R}^{M} \) is a shared MLP across users and iterations with a softmax output,  
yielding the refined symbol probability distribution for the next iteration.

\subsubsection{End-to-End Training and Loss Function}

Let the training dataset be denoted by $\{ (\tilde{\boldsymbol{y}}_t, \tilde{\boldsymbol{s}}_t) \}_{t=1}^{T}$, where $\tilde{\boldsymbol{y}}_t$ represents the received signal vector and $\tilde{\boldsymbol{s}}_t$ denotes the corresponding transmitted symbol vector for the $t$-th training sample. GNNSIC is trained by minimizing the average cross-entropy loss over all users:
\begin{align}
\mathcal{L}(\boldsymbol{\theta}) = \frac{1}{T K} \sum_{t=1}^{T} \sum_{i=1}^{K} -\log \left( \hat{\boldsymbol{p}}_{i}^{(L)}(\Tilde{\boldsymbol{y}}_{t}, \Tilde{\boldsymbol{h}}_{i,t}; \boldsymbol{\theta})_{\Tilde{s}_{i,t}} \right),
\end{align}
where \( \boldsymbol{\theta} \) denotes the collection of all model parameters. Here, the iteration index $l$ refers to repeated unrolling of the same
parameter-shared network block over the outer DAG.
This property will play a key role in the subsequent generalization analysis. The model is trained end-to-end using the Adam optimizer.

\subsubsection{Final Symbol Decision}

Upon completing the final iteration, the symbol detection for user \( i \) is given by $\hat{s}_i = \alpha_{m^*} \in \mathcal{M}^M$ such that
\begin{align}
m^* = \arg\max_{m \in \{1,2,\dots,M\} } \hat{\boldsymbol{p}}_{i,m}^{(L)}. 
\label{eq:hard_decision}
\end{align}
This rule selects the most probable symbol from the estimated posterior distribution.

\section{Generalization Analysis}
\label{sec:generalization_analysis}
Generalization analysis provides theoretical insight into why learning models can reliably perform well on unseen data, where a substantial body of work~\cite{Neyshabur2015Norm, Golowich2020Size, Bartlett2002Rademacher, Harvey2017Nearly, Bartlett2017Spectrally, Cao2019Generalization, Daniely2019Generalization, Wei2019Data}
has established norm-based generalization bounds for deep neural networks.
Motivated by these results, we develop a Rademacher complexity-based
generalization analysis for the \emph{Network-of-MLPs} function class that
explicitly accounts for its underlying DAG structure, and specialize the
resulting bounds to DeepSIC and GNNSIC as two structurally distinct
instantiations. This framework provides a
principled way to upper-bound the generalization gap, defined as the difference
between the expected test error and the empirical training error, for
multi-class classification models.

Rademacher complexity characterizes the richness of a hypothesis class by
quantifying its ability to correlate with random noise. Concretely, it measures
the extent to which functions in the class can fit random label assignments
generated by independent Rademacher variables taking values in $\{\pm1\}$.
A larger Rademacher complexity indicates a higher-capacity hypothesis class that
can more easily adapt to random fluctuations, which is commonly associated with
an increased risk of overfitting and a larger expected generalization
gap~\cite{Mohri2018Foundations, Shalev2014Understanding, Bartlett2002Rademacher}.

\textbf{Definition 1 (Empirical Rademacher Complexity).}
\label{def:1}
Let $\mathcal{W}$ denote the parameter space of the \emph{Network-of-MLPs}
architecture, and let
$\mathcal{F} \triangleq \{ f_{\boldsymbol{w}} : \boldsymbol{w} \in \mathcal{W} \}$
denote the associated hypothesis class.
Each $f_{\boldsymbol{w}} : \mathcal{X}^K \to \mathbb{R}^{K\times M}$ maps a joint 
$K$-block input to $K$ output vectors, each of dimension $M$.

Given a sample set
$\mathcal{S}_x = \{ \boldsymbol{x}_t \}_{t=1}^T$ with $\boldsymbol{x}_t \in \mathcal{X}^K$,
the empirical Rademacher complexity of $\mathcal{F}$ is defined as
\begin{align}
    \mathcal{R}_{\mathcal{S}_x}(\mathcal{F})
    :=
    \frac{1}{T}
    \mathbb{E}_{\xi}
    \left[
        \sup_{\boldsymbol{w} \in \mathcal{W}}
        \left|
            \sum_{t=1}^T
            \sum_{k=1}^K
            \sum_{m=1}^M
            \xi_{k,t,m}\,
            f_{\boldsymbol{w}}(\boldsymbol{x}_{t})_{k,m}
        \right|
    \right],
    \label{eq:rademacher_def}
\end{align}
where $\xi_{k,t,m}$ are independent Rademacher variables taking values
in $\{\pm1\}$.
This quantity measures the ability of $\mathcal{F}$ to fit random noise on the
finite sample set $\mathcal{S}_x$.

\vspace{0.5em}
We consider a multi-block classification task defined by a distribution
$P$ over $(\boldsymbol{x},\boldsymbol{s})\in \mathcal{X}^K \times [M]^K$.
For any $f_{\boldsymbol{w}}\in\mathcal{F}$, the expected classification error is
\begin{align}
\mathrm{err}_P(\boldsymbol{w})
\!:=\!
\frac{1}{K}\!
\sum_{k=1}^K\!
\mathbb{E}_{(\boldsymbol{x},\boldsymbol{s})\sim P}
\mathbb{I}\!\left[
\max_{m \neq s_k}
f_{\boldsymbol{w}}(\boldsymbol{x})_{k,m}
\!\ge\!
f_{\boldsymbol{w}}(\boldsymbol{x})_{k,s_k}
\right]\nonumber
\end{align}
where $\mathbb{I}(\cdot)$ is the indicator function.
Given independent and identically distributed (i.i.d.) training samples
$\mathcal{S}=\{(\boldsymbol{x}_t,\boldsymbol{s}_t)\}_{t=1}^T$,
the empirical $\gamma$-margin error is defined as
\begin{align}
\mathrm{err}_{S}^{\gamma} (\boldsymbol{w})
\!:=\!
\frac{1}{T}\!
\sum_{t=1}^T\!
\sum_{k=1}^K\!
\mathbb{I}\!\left[
\max_{m \neq s_{kt}}
f_{\boldsymbol{w}}(\boldsymbol{x}_{t})_{k,m}
\!+\! \gamma
\!\ge\!
f_{\boldsymbol{w}}(\boldsymbol{x}_{t})_{k,s_{kt}}
\right].\nonumber
\label{eq:margin_error}
\end{align}
\begin{customlemma}{1}[{\cite[Lemma~2.2]{galanti2023norm}}]\label{lemma:1}
With probability at least $1-\delta$ over the draw of ${\mathcal{S}}$, the following generalization error gap holds uniformly for all $f_{\boldsymbol{w}} \in \mathcal{F}$:
\begin{equation}
\mathrm{err}_P(\boldsymbol{w})
-
\mathrm{err}_{\mathcal{S}}^{\gamma}(\boldsymbol{w})
\;\le\;
\frac{2\sqrt{2}}{\gamma}\, \mathcal{R}_{\mathcal{S}_x}(\mathcal{F})
+
3\sqrt{\frac{\log(2/\delta)}{2T}}.
\label{eq:gap_bound}
\end{equation}
\end{customlemma}
\noindent Lemma~\ref{lemma:1} reduces the generalization analysis to controlling the empirical Rademacher complexity of the hypothesis class $\mathcal{F}$.
\subsection{Network Architecture Definitions}

We consider a multi-user predictor consisting of $K$ users interacting
through $L$ outer iterations. The model is composed of $K\times L$ network
blocks: for each user $k\in[K]$ and each outer iteration $l\in[L]$,
there exists a corresponding block $\mathcal{B}_{k_l}$.

Each block $\mathcal{B}_{k_l}$ is an inner MLP represented by a DAG of depth $N$, i.e., a feedforward computation graph with
directed edges and no cycles. At a fixed outer iteration $l$, the $K$ inner
blocks $(\mathcal{B}_{k_l})_{k=1}^{K}$ operate in parallel across users.
Across outer iterations, these blocks are coupled through learnable outer
weights, which induce an outer DAG of depth $L$ governing inter-user
information flow over iterations.

The resulting architecture thus forms a hierarchical
\emph{Network-of-MLPs}, obtained by composing inner DAGs through an outer DAG
with structured interactions across the $K$ network blocks.

\noindent\textbf{Inner MLP block $\mathcal{B}_{k_l}$.}
Fix a user $k\in[K]$ and an outer iteration $l\in[L]$.
Let $G^{(\mathrm{in})}_{k_l}=(\mathcal{V}^{(\mathrm{in})}_{k_l},\mathcal{E}^{(\mathrm{in})}_{k_l})$ denote the DAG associated
with the inner block $\mathcal{B}_{k_l}$, consisting of $N+1$ layers indexed
by $n=0,\dots,N$.
Following the notation in Section~\ref{sec:notation_network_of_mlp_arch}, we write
$\boldsymbol{z}^{(\mathrm{in}),n}_{k_l,j}$ for the $j$-th neuron in layer $n$ and
$\boldsymbol{v}^{(\mathrm{in}),n}_{k_l}$ for the collection of neurons in that layer.

The input to $\mathcal{B}_{k_l}$ is defined as
\begin{align*}
    \boldsymbol{x}_{k_l}
     =
    (\boldsymbol{z}^{(\mathrm{in}),0}_{k_l,j})_{j=1}^{q^{(\mathrm{in})}_0}.
\end{align*}
For inner layers $1\le n\le N-1$, the neuron activations follow the standard
MLP recursion
\begin{align*}
    \boldsymbol{z}^{(\mathrm{in}),n}_{k_l,j}(\boldsymbol{x}_{k_l})
    =
    \sigma\!\left(
        \boldsymbol{W}^{(\mathrm{in}),n}_{k_l,j}\,
        \boldsymbol{v}^{(\mathrm{in}),n-1}_{k_l,j}(\boldsymbol{x}_{k_l})
    \right),
\end{align*}
where $\sigma(\cdot)$ denotes the ReLU activation function,
$\boldsymbol{v}^{(\mathrm{in}),n-1}_{k_l,j}$ is the concatenation of predecessor activations,
and
$\boldsymbol{W}^{(\mathrm{in}),n}_{k_l,j}
\in
\mathbb{R}^{c_n\times (c_{n-1}\cdot|\mathrm{pred}(n,j)_{k_l}|)}$
is the corresponding weight matrix.

The inner block $\mathcal{B}_{k_l}$ produces an $M$-dimensional score vector
through a final linear readout layer,
\begin{align*}
    f^{(\mathrm{in})}_{k_l}(\boldsymbol{x}_{k_l})
    :=
    \boldsymbol{W}^{(\mathrm{in}),N}_{k_l}\,
    \boldsymbol{z}^{(\mathrm{in}),N-1}_{k_l}(\boldsymbol{x}_{k_l})
    \in \mathbb{R}^{M}.
\end{align*}
The sparsity of the inner block is characterized by the layer-wise in-degree
of its associated DAG.
Specifically, we quantify the inner connectivity at layer $n$ by the sparsity
degree $d^{(\mathrm{in})}_n$ and its worst-case counterpart
$\bar{d}^{(\mathrm{in})}_{l,n}$, as defined in
Section~\ref{sec:notation_network_of_mlp_arch}.

\noindent\textbf{Outer network (across $L$ steps and $K$ blocks).}
We consider the composition of $K$ network blocks over $L$ outer iterations.
The resulting outer architecture is represented by a DAG
$G^{(\mathrm{out})}=(\mathcal{V}^{(\mathrm{out})},\mathcal{E}^{(\mathrm{out})})$ with $L+1$ layers indexed by $l=0,\dots,L$.
Following the notation in Section~\ref{sec:notation_network_of_mlp_arch}, we denote by
$\boldsymbol{z}^{(\mathrm{out}),l}_{k_l}$ the $k_l$-th network block at outer layer $l$, and by
$\boldsymbol{v}^{(\mathrm{out}),l}$ the collection of network blocks at that layer.

The input to the outer network is defined as 
\begin{align*}
    \boldsymbol{x}
     =
(\boldsymbol{z}^{(\mathrm{out}),0}_{k_0})_{k=1}^{q^{(\mathrm{out})}_0},
\end{align*}
For outer layers $1\le l\le L$, the block-wise activations follow the recursive form
\begin{align*}
    \boldsymbol{z}^{(\mathrm{in}),0}_{k_l,j}
    &=
    \sigma\!\left(
        \boldsymbol{W}^{(\mathrm{out}),l}_{k_l}\,
        \boldsymbol{v}^{(\mathrm{out}),l-1}_{k_l}(\boldsymbol{x})
    \right),\\
    \boldsymbol{z}^{(\mathrm{out}),l}_{k_l}(\boldsymbol{x}) &= f^{(\mathrm{in})}_{k_l}(\boldsymbol{x}_{k_l}),
\end{align*}
where $\sigma(\cdot)$ denotes the ReLU activation function,
$\boldsymbol{v}^{(\mathrm{out}),l-1}_{k_l}$ denotes the concatenation of predecessor block activations,
and $\boldsymbol{W}^{(\mathrm{out}),l}_{k_l}$ is the corresponding aggregation weight matrix.

The sparsity of the outer composition is quantified by the layer-wise maximum
in-degree of the outer DAG, denoted by $d^{(\mathrm{out})}_l$
(see Section~\ref{sec:notation_network_of_mlp_arch}).

\noindent\textbf{Inner product-norm (per block).}
We define the inner product-norm, based on a product of maximum Frobenius norms, as
\begin{align}
    \rho_{\mathrm{inner}}(\boldsymbol{w}_{k_l})
    :=
    \|\boldsymbol{W}^{(\mathrm{in}),N}_{k_l}\|_{F}
    \cdot
    \prod_{n=1}^{N-1}
    \max_{i\in[q^{(\mathrm{in})}_n]}
    \|\boldsymbol{W}^{(\mathrm{in}),n}_{k_l,i}\|_{F},
    \label{eq:rho_inner_revised}
\end{align}
and assume a uniform bound
\(
\rho_{\mathrm{inner}}(\boldsymbol{w}_{k_l}) \le \bar{\rho}_{\mathrm{inner}}
\)
holds for all \(k\in[K]\) and \(l\in[L]\) throughout the analysis.

\noindent\textbf{Outer product-norm.}
We define the outer product-norm, based on a product of maximum Frobenius norms, as
\begin{align*}
    \rho_{\mathrm{outer}}(\boldsymbol{w})
    :=
    \|\boldsymbol{W}^{(\mathrm{out}),L}\|_{F}
    \cdot
    \prod_{l=1}^{L-1}
    \max_{k_l\in[q^{(\mathrm{out})}_l]}
    \|\boldsymbol{W}^{(\mathrm{out}),l}_{k_l}\|_{F},
\end{align*}
and assume a uniform bound
\(
\rho_{\mathrm{outer}}(\boldsymbol{w})\le \bar{\rho}_{\mathrm{outer}}
\)
holds for the outer composition throughout the analysis.

\noindent\textbf{Constrained hypothesis class.}
Given fixed inner- and outer-level capacity bounds
$\bar{\rho}_{\mathrm{inner}}>0$ and
$\bar{\rho}_{\mathrm{outer}}>0$,
we consider networks whose parameters satisfy
\begin{align*}
    \rho_{\mathrm{inner}}(\boldsymbol{w}_{k_l})
    \le \bar{\rho}_{\mathrm{inner}},
    \quad \forall k\in[K],\, l\in[L],
\end{align*}
and
\begin{align*}
    \rho_{\mathrm{outer}}(\boldsymbol{w})
    \le \bar{\rho}_{\mathrm{outer}}.
\end{align*}
These constraints induce a network-level complexity measure
\(
    \rho(\boldsymbol{w})
    := \Phi\bigl(
        \rho_{\mathrm{inner}}(\boldsymbol{w}_{k_l}),
        \rho_{\mathrm{outer}}(\boldsymbol{w})
    \bigr),
\) where $\Phi$ is a non-decreasing function in each
argument.

Let
\[
G
\triangleq
\bigl(
G^{(\mathrm{out})},
\{G^{(\mathrm{in})}_{k,l}\}_{k\in[K],\,l\in[L]}
\bigr)
\]
denote the overall computational DAG-of-DAGs of the \emph{Network-of-MLPs}.
We define the constrained hypothesis class as
\begin{align}
\mathcal{F}_{G,\rho}
:=
\bigl\{
f_{\boldsymbol{w}} \in \mathcal{F}_{G}
:
\rho(\boldsymbol{w}) \le \rho
\bigr\}.
\label{eq:rho}
\end{align}
The Rademacher complexity of the class $\mathcal{F}_{G,\rho}$, and the resulting
generalization bounds, are obtained via a two-level peeling argument,
consisting of an inner-level step that controls each block through
$\bar{\rho}_{\mathrm{inner}}$
and an outer-level step that captures the compositional
dependence across blocks through $\bar{\rho}_{\mathrm{outer}}$.
\subsection{Peeling Lemma for Network-of-MLPs}
The analysis relies on a standard peeling argument for one-layer neural networks,
originally established in~\cite{galanti2023norm}.
For completeness and to align with the notation used in this paper, we restate
the lemma below.
\begin{customlemma}{2}[Peeling Lemma {\cite[Lemma~3.4]{galanti2023norm}}]
\label{lemma:2}
For any fixed inputs $\{\boldsymbol{x}_{t}\}_{t=1}^T$, let $\sigma$ be a $1$-Lipschitz, positively homogeneous activation applied element-wise (e.g., ReLU).  
Consider a class of vector-valued functions 
\begin{align*}
    \mathcal{F}
    \subset
    \bigl\{
        f
        =
        (f_{1},\dots,f_{q})
        \,\big|\,
        f_{j} : \mathbb{R}^{a}\!\to\mathbb{R}^{p},\; j\in[q]
    \bigr\},
\end{align*}
and let $g:\mathbb{R}\to[0,\infty)$ be any convex and monotonically increasing function.  
Then, for any radius $R>0$, the following inequality holds:
\begin{align}
\label{eq:peeling_lemma}
    &\mathbb{E}_\xi 
    \sup_{\substack{
    f\in \mathcal{F},\;\\
        \boldsymbol{W}:\|\boldsymbol{W}_{j}\|_{F}\le R
    }}
    g\!\Biggl(
            \sqrt{
            \sum_{j=1}^{q}
            \Bigl\|
                \sum_{t=1}^{T}
                \xi_{t}\,
                \sigma\!\bigl(
                    \boldsymbol{W}_{j} f_{j}(\boldsymbol{x}_{t})
                \bigr)
            \Bigr\|_{2}^{2}
            }
    \Biggr)
    \nonumber\\[1mm]
    &\quad\le
    2\,
    \mathbb{E}_{\xi}\!
    \left[
        \sup_{\substack{f\in \mathcal{F}, \\j\in[q]}}
        % \sup_{j\in[q]}
        g\!\left(
            \sqrt{q}\, R\,
            \Bigl\|
                \sum_{t=1}^{T}
                \xi_{t}\,
                f_{j}(\boldsymbol{x}_{t})
            \Bigr\|_{2}
        \right)
    \right],
\end{align}
where the supremum on the left is taken over all weight matrices $\{\boldsymbol{W}_{j}\}_{j=1}^{q}$ satisfying $\|\boldsymbol{W}_{j}\|_{F}\le R$.
\begin{proof}
See Appendix~\ref{app:4.3}.
\end{proof}
\end{customlemma}
\subsection{Theoretical Analysis}
We now establish an upper bound on the empirical Rademacher complexity
of the constrained hypothesis class $\mathcal{F}_{G,\rho}$
associated with the \emph{Network-of-MLPs} architecture.

\begin{customprop}{1}
\label{prop:1}
For a fixed sample set
$\mathcal{S}_x=\{\boldsymbol{x}_t\}_{t=1}^T \subset \mathcal{X}^K$,
the following bound holds.
Then, the empirical Rademacher complexity of $\mathcal{F}_{G,\rho}$
admits the following upper bound:
\begin{align}
    \mathcal{R}_{\mathcal{S}_x}(\mathcal{F}_{G,\rho})
    \;\le\;
    \frac{K\bar{\rho}_{\mathrm{inner}}^{\,L}\,\bar{\rho}_{\mathrm{outer}}\,}{T}
    (1+\sqrt{2A})\sqrt{B},
    \label{eq:prop_result}
\end{align}
where the quantities $A$ and $B$ are defined as
\begin{align*}
A &\coloneqq (N+1)L\log 2
+ L\log M
\\
&\qquad+ \sum_{l=1}^{L}\log d^{(\mathrm{out})}_l
+ \sum_{l=1}^{L}\sum_{n=1}^{N-1}\log d^{(\mathrm{in})}_{l,n},\\
B &\coloneqq
\max_{k_{0},\dots,k_L}
\prod_{l=1}^{L} |\mathrm{pred}(l,k_l)|
\max_{j_0,\dots,j_N}
\prod_{l=1}^{L}\prod_{n=1}^{N-1}
|\mathrm{pred}(n,j_n)_{k_l}| \\
&\qquad\cdot
\max_{k_{0}}
\sum_{t=1}^{T}
\big\|
\boldsymbol{z}^{(\mathrm{out}),0}_{k_{0}}(\boldsymbol{x}_t)
\big\|_2^2.
\end{align*}
The maximization in $B$ is taken over all indices corresponding to valid
paths in the network, i.e., such that
$j_{n-1} \in \mathrm{pred}(n,j_n)_{k_l}$ for all $n$ and $l$,
and $k_{l-1} \in \mathrm{pred}(l,k_l)$.
Here, $M$ denotes the output dimension.
\end{customprop}

\begin{proof}
See Appendix~\ref{app:prop 4.4}.
\end{proof}
\begin{customthm}{1}
\label{thm:1}
Let $\mathcal{F}_{G,\rho}$ be the constrained hypothesis class defined in
\eqref{eq:rho}, with
\(
\rho = \bar\rho_{\mathrm{inner}}^{\,L}\bar\rho_{\mathrm{outer}}.
\)
Let $P$ be a distribution over $\mathcal{X}^{K} \times [M]^{K}$, and let
${\mathcal{S}} = \{(\boldsymbol{x}_t, \boldsymbol{s}_t)\}_{t=1}^{T}$ be a dataset of $T$ i.i.d.\
samples drawn from $P$.
Then, for any margin parameter $\gamma>0$, with probability at least
$1-\delta$ over the draw of ${\mathcal{S}}$, the following generalization bound holds
uniformly for all $f_{\boldsymbol{w}} \in \mathcal{F}_{G,\rho}$:
\begin{align}
    \mathrm{err}_P(\boldsymbol{w}) - \mathrm{err}_{\mathcal{S}}^\gamma(\boldsymbol{w})
    \;\le\;&
    \frac{2\sqrt{2}\,K\,(\rho+1)}{\gamma T}
    \bigl(1+\sqrt{2A}\bigr)\sqrt{B} \nonumber\\
    &\;+\;
    3\sqrt{\frac{\log\!\bigl(2(\rho+2)^2/\delta\bigr)}{2T}},
\label{eq:final_thm_bound}
\end{align}
where $A$ and $B$ are the architecture- and data-dependent complexity
terms defined in Proposition~\ref{prop:1}.
\end{customthm}

\begin{proof}
See Appendix~\ref{app:The 4.5}.
\end{proof}
\begin{customcor}{1}
\label{cor:1}
Consider the setting of Proposition~\ref{prop:1} and Theorem~\ref{thm:1}.
Assume further that the interaction between network blocks across iterations
is implemented via feature concatenation, and that no additional learnable
linear transformation is applied between blocks.
Then, the induced outer operator has unit norm, and the network-level norm bound simplifies to
\begin{align*}
\rho = \bar\rho_{\mathrm{inner}}^{\,L}.
\end{align*}
\end{customcor}
\noindent\textbf{Comparison of the Generalization Bounds: }
We compare the theoretical generalization bounds of DeepSIC and GNNSIC to 
highlight the impact of parameter sharing. 

\paragraph*{DeepSIC}
The DeepSIC architecture follows the standard \emph{Network-of-MLPs} architecture which consists of $K$ distinct neural network blocks
unrolled over $L$ iterations.
The input feature vector for the $k$-th user at time index $t$, denoted as
$\boldsymbol{x}_{t,k}^{\mathrm{Deep}} \in \mathbb{R}^{N+(K-1)(M-1)}$,
is constructed by concatenating the received signal
$\boldsymbol{y}_t \in \mathbb{R}^{N}$ with the soft symbol estimates from the
previous iteration $(l-1)$ for the remaining $K-1$ users,
$(\hat{\boldsymbol{p}}_{t,i}^{(l-1)})_{i\neq k}$.
Since the interaction between blocks is implemented solely via feature
concatenation, Corollary~\ref{cor:1} applies directly.

\smallskip
With the notation and assumptions in
Section~\ref{sec:notation_network_of_mlp_arch},
the Rademacher complexity of DeepSIC is governed by the auxiliary terms
$A_{\mathrm{D}}$ and $B_{\mathrm{D}}$, which admit the bounds
\begin{align*}
A_{\mathrm{D}}
\;&\le\;
L
\log\!\Big(
2^{N+1}
M
\bar{d}^{(\mathrm{out})}
(\bar{d}^{(\mathrm{in})})^{N-1}
\Big),\\
B_{\mathrm{D}}
\;&\le\;
(\bar{d}^{(\mathrm{out})})^L
(\bar{d}^{(\mathrm{in})})^{L(N-1)}
\cdot E_0 .
\end{align*}

The resulting generalization gap is therefore bounded by
\begin{align}
\label{eq:final_gen_bound_DeepSIC}
\tilde{\mathcal{O}}\!\left(
\frac{K(\rho+1)}{T}
(\bar{d}^{(\mathrm{out})})^{\frac{L}{2}}
(\bar{d}^{(\mathrm{in})})^{\frac{L(N-1)}{2}}
\sqrt{L E_0}
\right),
\end{align}
where $\rho=\bar\rho_{\mathrm{inner}}^{\,L}$ follows from
Corollary~\ref{cor:1}.

\paragraph*{GNNSIC}
GNNSIC can be viewed as a special case of
the proposed \emph{Network-of-MLPs} framework with a single outer layer ($L=1$)
and fully shared parameters. GNNSIC substantially reduces the hypothesis space complexity
through two forms of parameter sharing:
\emph{(i) cross-user sharing}, whereby all users employ a single shared neural
network block, and
\emph{(ii) cross-iteration sharing}, whereby the same parameters are reused
across message-passing iterations.
As a result, the depth-dependent complexity terms in the generalization bound
effectively collapse to $L=1$, and the corresponding degrees are independent of
the outer iteration index. Moreover, due to cross-user sharing, the maximization
over computation paths reduces to a single user index. 
The input feature
$\boldsymbol{x}_{t,k}^{\mathrm{GNN}} \in \mathbb{R}^{2N+M}$
consists of the received signal $\boldsymbol{y}_t$, the channel vector
$\boldsymbol{h}_{kt}$, and the soft symbol estimate
$\hat{\boldsymbol{p}}_{kt}^{(l-1)}$.
Consequently, under the global notation introduced in 
Section~\ref{sec:notation_network_of_mlp_arch}, the auxiliary complexity terms for GNNSIC
can be expressed as
\begin{align*}
A_{\mathrm{G}}
\;&\le\;
\log\!\Big(
2^{N+1}
M
\bar{d}^{(\mathrm{out})}
(\bar{d}^{(\mathrm{in})})^{N-1}
\Big),\\
B_{\mathrm{G}}
\;&\le\;
\bar{d}^{(\mathrm{out})}
(\bar{d}^{(\mathrm{in})})^{N-1}
\cdot E_0 ,
\end{align*}
where $E_0$ is the input energy bound defined in
Section~\ref{sec:notation_network_of_mlp_arch}.
The resulting generalization gap for GNNSIC is therefore bounded by
\begin{align}
\tilde{\mathcal{O}}\!\left(
\frac{(\rho'+1)}{T}
(\bar{d}^{(\mathrm{out})})^{\frac{1}{2}}
(\bar{d}^{(\mathrm{in})})^{\frac{N-1}{2}}
\sqrt{
E_0
}
\right),
\end{align}
where $\rho'=\bar\rho_{\mathrm{inner}}$ due to parameter sharing. Compared with DeepSIC, the absence of depth-dependent growth in both
$A_{\mathrm{G}}$ and $B_{\mathrm{G}}$ highlights the crucial role of parameter sharing in controlling the hypothesis space complexity.

The derived Rademacher complexity bounds expose a fundamental structural
difference between DeepSIC and the proposed GNNSIC. For DeepSIC, the dominant
terms in the bound grow exponentially with the number of SIC iterations, as a
direct consequence of norm accumulation and degree amplification induced by
repeatedly composing non-shared neural modules along the outer DAG. In contrast,
GNNSIC avoids this exponential growth by enforcing structured parameter sharing
across both users and iterations, which effectively collapses the outer DAG and
eliminates both effects. From a system design perspective, this analysis suggests
that GNNSIC is particularly well suited for systems with a large number of users
or deeper SIC iterations, where the exponential growth in model complexity
inherent to conventional DeepSIC architectures may impose fundamental
generalization limitations. By decoupling the generalization behavior from the
number of SIC iterations, GNNSIC provides a theoretically grounded, more robust
and generalizable framework for MIMO detection.
\section{Numerical Results}
\label{sec:numerical_results}

This section presents the performance evaluation of the proposed GNNSIC model compared with the baseline DeepSIC approach. For $6\times 6$ case, all models are trained on $48{,}000$ samples using the ADAM optimizer with a learning rate of $0.005$, and validated on $12{,}000$ samples. To ensure stable SER estimates, the number of testing samples increases with the Signal-to-Noise Ratio (SNR). Specifically, for the linear synthetic channel with $\mathrm{SNR} = 0$ to $10$ dB, $20{,}000$ samples are used; in higher SNR regime, the number increases to $200{,}000$ at $12$ dB and $2{,}000{,}000$ at $14$ dB. Training is conducted over $100$ epochs with a batch size of $64$, while validation and testing use a batch size of $128$. 
Binary phase-shift keying (BPSK) modulation is considered for simplicity, with $\mathcal{S} = \{-1, +1\}$ and $M = 2$.

\subsection{Linear Synthetic Channels}
\label{sub:scenario1}

We first evaluate the proposed GNNSIC on a linear synthetic Gaussian channel, where the channel model and the CSI uncertainty follow exactly the settings for DeepSIC in~\cite{shlezinger2020deepsic}. 
In particular, we use \textit{perfect CSI} to indicate training and testing on the same channel realization, while \textit{CSI uncertainty} means training on a perturbed channel $\hat{h}=h+e$ with $e \sim \mathcal{N}(0, \sigma_e^2)$ and testing on the unperturbed channel $h$.
All learning-based approaches are trained at 10 dB and tested at other SNRs for generalization evaluation.

\begin{figure}[!t]
    \centering
    \includegraphics[width=\linewidth]{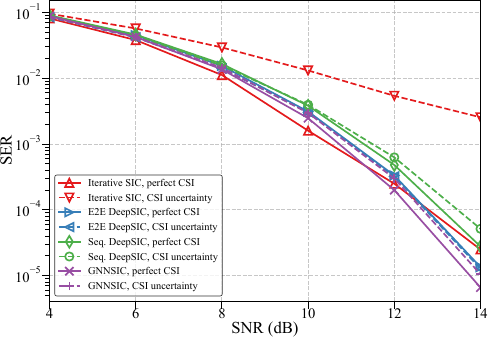}
    \caption{SER performance versus SNR for GNNSIC, iterative SIC, and DeepSIC over a $6 \times 6$ linear Gaussian channel with perfect CSI ($\sigma_e^2 = 0$) and CSI uncertainty ($\sigma_e^2 = 0.1$). }
    \label{fig:linear-gaussian-label-6}
\end{figure}

As shown in Fig.~\ref{fig:linear-gaussian-label-6}, under perfect CSI, GNNSIC achieves performance comparable to that of \textit{E2E DeepSIC}, and outperforms both iterative SIC and \textit{Seq. DeepSIC}, demonstrating its ability to approximate DeepSIC’s inference behavior within a structured message-passing framework. Within the entire SNR regime, GNNSIC closely follows the SER trajectory of \textit{E2E DeepSIC}, with certain performance improvements at high SNRs. Under CSI uncertainty, both GNNSIC and DeepSIC exhibit strong robustness against channel perturbations, significantly outperforming the iterative SIC baseline. Notably, GNNSIC consistently yields lower SERs than \textit{E2E DeepSIC} at all SNRs with or without CSI uncertainty, highlighting the enhanced generalization capability and robustness of GNNSIC. 

\subsection{Non-Linear Synthetic Channels}
\label{sub:scenario2}

In this experiment, we evaluate the performance of GNNSIC on a non-linear synthetic channel against the baselines. This setting aims to demonstrate that GNNSIC can achieve comparable performance to DeepSIC, even in quantized and Poisson communication scenarios, which are given in~\cite{shlezinger2020deepsic}.
Since the sequential training variant of DeepSIC (i.e., \textit{Seq. DeepSIC}) adopts the same network architecture as the end-to-end trained version (\textit{E2E DeepSIC}) but exhibits inferior performance on synthetic linear Gaussian channels, we omit \textit{Seq. DeepSIC} in subsequent comparisons and focus solely on \textit{E2E DeepSIC}.

\begin{figure}[!htbp]
    \centering
    \includegraphics[width=\linewidth]{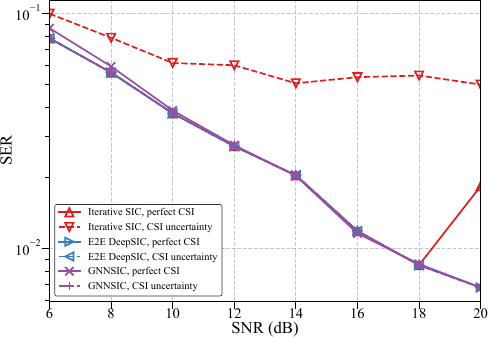}
    \caption{SER versus SNR for GNNSIC, iterative SIC, and DeepSIC over a $4 \times 4$ synthetic quantized Gaussian channel with \textit{perfect CSI} ($\sigma_e^2 = 0$) and \textit{CSI uncertainty} ($\sigma_e^2 = 0.1$).}
    \label{fig:quantized-gaussian-channel-label}
\end{figure}

\begin{figure}[!htbp]
    \centering
    \includegraphics[width=\linewidth]{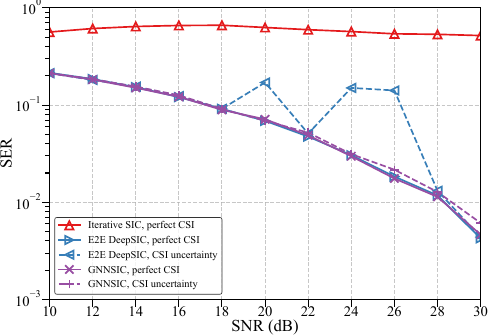}
    \caption{SER versus SNR for GNNSIC, iterative SIC, and DeepSIC over a $4 \times 4$ synthetic Poisson channel with \textit{perfect CSI}  ($\sigma_e^2 = 0$) and \textit{CSI uncertainty}  ($\sigma_e^2 = 0.1$).}
    \label{fig:poisson-channel}
\end{figure}

As illustrated in Fig.~\ref{fig:quantized-gaussian-channel-label}, GNNSIC achieves performance comparable to that of \textit{E2E DeepSIC} under both perfect and imperfect CSI conditions in the quantized Gaussian channels. This indicates that GNNSIC is capable of effective symbol detection even in the presence of significant nonlinear distortions. 
The results for the Poisson channel are presented in Fig.~\ref{fig:poisson-channel}, where GNNSIC achieves the lowest SER in all SNR regimes, significantly outperforming iterative SIC in the perfect CSI scenario. This highlights its superior capability in mitigating multiuser interference and enhancing detection accuracy. Moreover, under CSI uncertainty, GNNSIC maintains a notable performance advantage, exhibiting reduced degradation compared to both iterative SIC and \textit{E2E DeepSIC}. These results demonstrate the robustness of GNNSIC against channel estimation errors.

\subsection{Robustness to CSI Uncertainty}
\label{sub:scenario5}

\begin{figure}[!htbp]
    \centering
    \includegraphics[width=\linewidth]{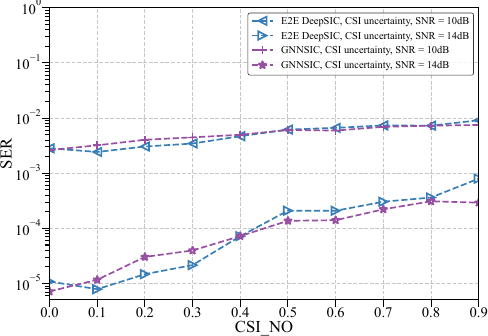}
    \caption{SER versus CSI\_NO for GNNSIC and DeepSIC at $\mathrm{SNR} = 10\,\mathrm{dB}$ and $14\,\mathrm{dB}$ on a $6 \times 6$ synthetic linear Gaussian channel.}
    \label{fig:CSI_NO}
\end{figure}

To evaluate the robustness of GNNSIC and \textit{E2E DeepSIC} under CSI uncertainty, we conduct experiments with varying levels of CSI noise (denoted as CSI\_NO) at two fixed SNRs: $10\,\mathrm{dB}$ and $14\,\mathrm{dB}$.
As illustrated in Fig.~\ref{fig:CSI_NO}, GNNSIC maintains as stable SER performance as DeepSIC when CSI uncertainty increases. At $10\,\mathrm{dB}$ SNR, the SER of both GNNSIC and DeepSIC increases slightly with CSI\_NO. 
The effect of CSI uncertainty becomes more prominent at $14\,\mathrm{dB}$, where both GNNSIC and DeepSIC experience significant performance degradation when CSI uncertainty increases. Notably, GNNSIC exhibits a plateau in SER when CSI\_NO values are high, indicating stronger resilience to severe CSI uncertainty. 

\subsection{User Generalization}
\label{sub:scenario4}
Thanks to the advantage of cross-user parameter sharing, the proposed GNNSIC possesses excellent user generalization capability, that is, the trained GNNSIC model for a large number of users can be directly deployed to the MIMO scenarios with a small number of users \textit{without re-training}. This is beyond the capability of DeepSIC, where the models should be \textit{retrained} again and again for different user configurations.

This subsection investigates the user generalization capability of the proposed GNNSIC model in multiuser MIMO systems when the number of transmit users ($K$) varies. The GNNSIC detector is trained on a fixed $32 \times 32$ MIMO configuration, and is evaluated under different numbers of users, specifically $K = 8, 16, 32$. Comparative results are presented against both iterative SIC and DeepSIC baselines.

\begin{figure}[!htbp]
    \centering
    \includegraphics[width=\linewidth]{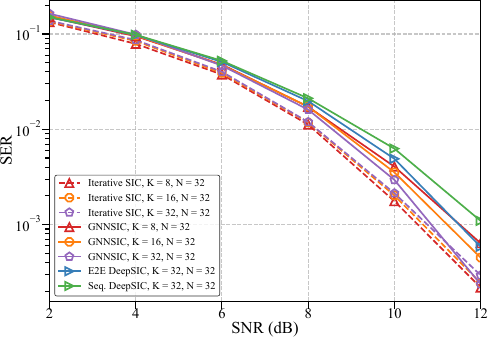}
    \caption{User generalization performance of GNNSIC in terms of SER versus SNR on a synthetic linear Gaussian MIMO channel. The GNNSIC model is trained with $K=32$ users and $N=32$ antennas, and evaluated under different numbers of users ($K=8, 16, 32$). Results are compared with the iterative SIC algorithm under matching $K$ values. Additionally, the performance of DeepSIC under the $32 \times 32$ MIMO setting is provided for reference.}
    \label{fig:user-generalization}
\end{figure}

As illustrated in Fig.~\ref{fig:user-generalization}, the proposed GNNSIC model demonstrates excellent user generalization capability, maintaining consistent SER performance across different user configurations. For $K = 8$, $16$, and $32$, GNNSIC achieves similar performance levels at low to moderate SNRs. As the number of users increases from $K = 8$ to $K = 32$, the performance gap between GNNSIC and iterative SIC gradually narrows. Notably, at $K = 32$, GNNSIC significantly outperforms DeepSIC and closely matches the performance of iterative SIC, even surpassing it in the high SNR regime (e.g., at $\mathrm{SNR} = 12$~dB). 
\subsection{Sample Efficiency}
\label{sub:scenario6}

\begin{table}[t]
\centering
\caption{Numerical Comparison of Rademacher Complexity Terms for DeepSIC and GNNSIC over $6\times6$ MIMO.}
\label{tab:complexity_comparison}
\renewcommand{\arraystretch}{1.25}
\setlength{\tabcolsep}{3pt}
\footnotesize
\begin{tabular}{lcc}
\hline
\hline
\textbf{Quantity} & \textbf{DeepSIC} & \textbf{GNNSIC} \\
\hline
Number of blocks $K$ & $6$ & $1$ (shared) \\
Iterations & $L = 5$ & $L=1$ (shared) \\
MLP depth $N$ & $2$ & $5$ \\
Constellation size $M$ & $2$ & $2$ \\
$ d^{(\mathrm{out})}_{l}, d^{(\mathrm{in})}_{l,n}$ & $(1),(30,60)$ & $(1),(14,28,2,15,28)$ \\
\hline
$\rho$ & $\bar\rho_{\mathrm{inner}}^{\,L}$ & $\bar\rho_{\mathrm{inner}}$ \\
\hline
Generalization gap &
$\mathcal{O}\!\left( \frac{K(\rho+1)}{T} \sqrt{A_{\mathrm{D}}B_{\mathrm{D}}} \right)$ &
$\mathcal{O}\!\left( \frac{(\rho+1)}{T} \sqrt{A_{\mathrm{G}}B_{\mathrm{G}}} \right)$ \\
\hline
$T$ for gap $\epsilon=0.1$
 & $\sim 10^{7}$ 
 & $\sim 10^{4}$ \\
\hline
\hline
\end{tabular}
\end{table}

The complexity comparison between DeepSIC and GNNSIC is summarized in
Table~\ref{tab:complexity_comparison}. Due to the cross-user and cross-iteration
parameter sharing, GNNSIC admits a substantially smaller hypothesis class,
leading to several orders-of-magnitude reduction in the worst-case
generalization bound compared with DeepSIC.

It is important to emphasize that such a pronounced theoretical gap does not
imply equally large differences in the empirical SER curves. This discrepancy is
expected, since the derived bounds characterize worst-case deviations over the
entire hypothesis class, whereas the trained networks correspond to specific
low-norm solutions favored by stochastic first-order optimization under
structured communication data. In the considered $6\times6$ MIMO setting with
i.i.d.\ Gaussian channels and BPSK signaling, the underlying detection task is
relatively simple, allowing even overparameterized architectures such as
DeepSIC to generalize well when trained with sufficient data.

\begin{figure}[!t]
    \centering
    \includegraphics[width=\linewidth]{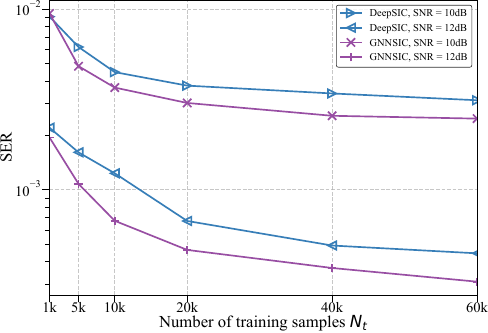}
    \caption{SER versus the number of training samples for GNNSIC and DeepSIC at
    $\mathrm{SNR} = 10\,\mathrm{dB}$ and $12\,\mathrm{dB}$ over a $6 \times 6$
    synthetic linear Gaussian channel.}
    \label{fig:training_samples}
\end{figure}

To further elucidate the improved sample efficiency observed in
Fig.~\ref{fig:training_samples}, Table~\ref{tab:complexity} reports the number of
trainable parameters of DeepSIC and GNNSIC under different MIMO configurations.
Across all settings, GNNSIC consistently requires one to two orders of magnitude
fewer parameters than DeepSIC, while preserving comparable expressivity through
structured parameter sharing.

\begin{table}[t]
\centering
\caption{Trainable Parameters of Different Detectors}
\label{tab:complexity}
\renewcommand{\arraystretch}{1.2}
\begin{tabular}{lcc}
\toprule
\textbf{MIMO Setting} & \textbf{DeepSIC} & \textbf{GNNSIC} \\
\midrule
$6\times 6$   & $25{,}260$   & $1{,}061$ \\
$32\times 32$ & $633{,}920$  & $18{,}689$ \\
$64\times 64$ & $2{,}496{,}640$ & $70{,}081$ \\
\bottomrule
\end{tabular}
\end{table}

The substantially reduced model complexity translates into superior sample
efficiency in practice. As shown in Fig.~\ref{fig:training_samples}, GNNSIC
achieves a target SER with significantly fewer training samples than DeepSIC at
both $\mathrm{SNR}=10\,\mathrm{dB}$ and $12\,\mathrm{dB}$. For instance, at
$10\,\mathrm{dB}$, GNNSIC attains an SER of approximately $3.5\times10^{-3}$ using
only $10$k training samples, whereas DeepSIC requires around $60$k samples to
reach comparable performance. Similar trends are observed at higher SNRs. These
results confirm that the proposed architecture improves generalization
by effectively controlling model complexity, rather than merely optimizing
average-case performance.

\section{Conclusion}
\label{sec:conclusion}

This paper studied fully data-driven MU-MIMO detection within a unified
\emph{Network-of-MLPs} framework, which models learning-based receivers as
networks of MLP blocks interconnected by outer and inner DAGs. Within this
framework, we reformulated soft interference cancellation as a graph-based
message-passing process and proposed GNNSIC, which exploits parameter sharing
across users and iterations. We further developed a norm-based generalization analysis for \emph{Network-of-MLPs} using
Rademacher complexity and DAG peeling, which reveals an exponential dependence of the
generalization bound on the SIC depth for DeepSIC, eliminated in GNNSIC through
parameter sharing. Simulation results demonstrate that GNNSIC attains comparable
or improved SER performance with fewer parameters and training samples, while
generalizing effectively across different user configurations.

\appendix
\subsection{Proof of Lemma~\ref{lemma:2}}
\label{app:4.3}
For self-containedness, we provide a proof of Lemma~\ref{lemma:2}, following the arguments
of~\cite{galanti2023norm}. This lemma will be repeatedly instantiated in both the inner-level and
outer-level peeling arguments developed in the subsequent sections.

Let $\boldsymbol{W}_{j}\in\mathbb{R}^{b\times p}$ denote its weight matrix, where $b$ corresponds to the output dimension in the present setting,
and write $\boldsymbol{w}_{j,1}^{\top},\dots,\boldsymbol{w}_{j,b}^{\top}$ for its rows associated with the $j$-th output neuron. For notational convenience, define for any $\boldsymbol{a}\in\mathbb{R}^p$ the scalar functional
\begin{align*}
Q_{j}(\boldsymbol{a})
\triangleq
\left(
    \sum_{t=1}^T \xi_{t}\,
    \sigma\!\left(
        \frac{\boldsymbol{a}^\top}{\|\boldsymbol{a}\|_2}\,
        f_{j}(\boldsymbol{x}_{t})
    \right)
\right)^2.
\end{align*}
With this definition, a direct expansion yields
\begin{align*}
&\sum_{j=1}^q
\left\|
    \sum_{t=1}^T
    \xi_{t}\,
    \sigma\!\left(
        \boldsymbol{W}_{j} f_{j}(\boldsymbol{x}_{t})
    \right)
\right\|_2^2 \\
&\quad=
\sum_{j=1}^q \sum_{\mu=1}^b
\|\boldsymbol{w}_{j,\mu}\|_2^2
\Bigg(
    \sum_{t=1}^T
    \xi_{t}\cdot\sigma\left(
        \frac{\boldsymbol{w}_{j,\mu}^\top}{\|\boldsymbol{w}_{j,\mu}\|_2}
        f_{j}(\boldsymbol{x}_{t})
    \right)
\Bigg)^2 \\
&\quad=
\sum_{j=1}^q \sum_{\mu=1}^b
\|\boldsymbol{w}_{j,\mu}\|_2^2\,
Q_{j}(\boldsymbol{w}_{j,\mu}) .
\end{align*}
For any collection $(\boldsymbol{w}_{j,\mu})_{\mu=1}^b$ satisfying $\|\boldsymbol{W}_{j}\|_F\leq R$, we clearly have
\begin{align}\label{eq:lemma43_max}
\sum_{\mu=1}^b
\|\boldsymbol{w}_{j,\mu}\|_2^2\,
Q_{j}(\boldsymbol{w}_{j,\mu})
\;\leq\;
R^2 \max_{\mu\in[b]} Q_{j}(\boldsymbol{w}_{j,\mu}) .
\end{align}
The upper bound is attained by concentrating the entire Frobenius norm on a single row, namely by choosing $\hat {\boldsymbol{w}}_{j,\mu^\star}$ with $\|\hat{\boldsymbol{w}}_{j,\mu^\star}\|_2=R$ for some $\mu^\star\in[b]$ and setting all remaining rows to zero. Since $g$ is monotone increasing, this implies
\begin{align}
&\mathbb{E}_{\xi}
\sup_{\substack{
    f\in \mathcal{F} \\
    \boldsymbol{W}_{j}:\,\|\boldsymbol{W}_{j}\|_F\leq R
}}
g\!\Bigg( \sqrt{ \sum_{j=1}^q \Bigg\| \sum_{t=1}^T \xi_{t}  \cdot\;\sigma(\boldsymbol{W}_{j} f_{j}(\boldsymbol{x}_{t})) \Bigg\|_2^2 } \Bigg)  \nonumber\\
&\quad\labelle{a}
\mathbb{E}_{\xi}
\sup_{\substack{
    f\in \mathcal{F} \\
    \|\boldsymbol{w}_{j}\|_2=R,\; j\in[q]
}}
g\!\Bigg(\sqrt{
        \sum_{j=1}^q
        \Bigg|
            \sum_{t=1}^T
            \xi_{t}\cdot\;\sigma\!\left(
                \boldsymbol{w}_{j}^\top
                f_{j}(\boldsymbol{x}_{t})
            \right)
        \Bigg|^2
    }
\Bigg)\nonumber\\
&\quad\labelle{b}
\mathbb{E}_{\xi}
\sup_{\substack{
    j\in[q],\;
    f\in \mathcal{F},\;
    \|\boldsymbol{w}\|_2=R
}}
g\!\Bigg(
    \sqrt{q}\,
    \Bigg|
        \sum_{t=1}^T
        \xi_{t}\cdot\;\sigma\!\left(
            \boldsymbol{w}^\top
            f_{j}(\boldsymbol{x}_{t})
        \right)
    \Bigg|
\Bigg)\nonumber\\
&\quad\labelle{c}
2\,
\mathbb{E}_{\xi}
\sup_{\substack{
    j\in[q],\;
    f\in \mathcal{F},\;
    \|\boldsymbol{w}\|_2=R
}}
g\!\Bigg(
    \sqrt{q}
    \sum_{t=1}^T
    \xi_{t}\cdot\;\sigma\!\left(
        \boldsymbol{w}^\top
        f_{j}(\boldsymbol{x}_{t})
    \right)
\Bigg)\nonumber\\
&\quad\labelle{d}
2\,\mathbb{E}_{\xi}
\sup_{\substack{
    j\in[q],\;
    f\in \mathcal{F}
}}
g\!\left(
    \sqrt{q}\,R\,
    \left\|
        \sum_{t=1}^T
        \xi_{t}\,
        f_{j}(\boldsymbol{x}_{t})
    \right\|_2
\right),
\label{eq:proof_peeling_lemma}
\end{align}
where $(b)$ bounds the sum by its largest term,
$(c)$ uses the inequality $g(|z|)\le g(z)+g(-z)$ together with the symmetry of the
Rademacher variables $\{\xi_t\}$,
and $(d)$ follows from Equation~(4.20) in~\cite{Ledoux1991Probability} and the
Cauchy--Schwarz inequality.
This completes the proof. \hfill \IEEEQED

\subsection{Proof of Proposition~\ref{prop:1}}
\label{app:prop 4.4}
Fix an arbitrary function $f_{\boldsymbol w} \in \mathcal{F}_{G,\rho}$, with $\rho$
defined in~\eqref{eq:rho}. Let
$\boldsymbol{W}^{(\mathrm{in}),n}_{k_l,j_n}\in\mathbb{R}^{c_n\times
(c_{n-1}|\mathrm{pred}(n,j_n)_{k_l}|)}$
denote the weight matrix of the $j_n$-th channel-generating unit in layer $n$
of the inner MLP for user $k\in[K]$ at outer iteration $l\in[L]$. Exploiting the positive homogeneity of ReLU, $\sigma(\alpha x)=\alpha\sigma(x)$,
any $f_{\boldsymbol w}\in\mathcal{F}_{G,\rho}$ admits an equivalent
re-parameterization $f_{\hat{\boldsymbol w}}$ obtained by layerwise normalization.

For the inner MLPs, we normalize the output layer of the $k$-th block at iteration $l$ as
\begin{align*}
\hat{\boldsymbol{W}}^{(\mathrm{in}),N}_{k_l}
\!\!&:=\!\!
\bar{\rho}_{\mathrm{inner}}
\frac{\boldsymbol{W}^{(\mathrm{in}),N}_{k_l}}{\|\boldsymbol{W}^{(\mathrm{in}),N}_{k_l}\|_F},\\
\hat{\boldsymbol{W}}^{(\mathrm{in}),n}_{k_l,j_n}
\!\!&:=\!\!
\frac{\boldsymbol{W}^{(\mathrm{in}),n}_{k_l,j_n}}{\max_{j_n}\|\boldsymbol{W}^{(\mathrm{in}),n}_{k_l,j_n}\|_F},
 n<N.
\end{align*}
Similarly, for the outer network,
\begin{align*}
\hat{\boldsymbol{W}}^{(\mathrm{out}),L}
\!\!&:=\!\!
\bar{\rho}_{\mathrm{outer}}
\frac{\boldsymbol{W}^{(\mathrm{out}), L}}{\|\boldsymbol{W}^{(\mathrm{out}), L}\|_F},\\
\hat{\boldsymbol{W}}^{(\mathrm{out}), l}_{j_l}
\!\!&:=\!\!
\frac{\boldsymbol{W}^{(\mathrm{out}),l}_{j_l}}{\max_{j_l}\|\boldsymbol{W}^{(\mathrm{out}),l}_{j_l}\|_F},
\; l<L.
\end{align*}
By construction and the positive homogeneity of ReLU, the re-parameterized network
$f_{\hat{\boldsymbol w}}$ realizes the same input--output mapping as $f_{\boldsymbol w}$, and hence
\begin{align*}
\mathcal{F}_{G,\rho}
\;\subseteq\;
\hat{\mathcal{F}}_{G,\rho},
\end{align*}
where
\begin{align*}
\hat{\mathcal{F}}_{G,\rho}
\;:=
\Bigl\{
f_{\hat{\boldsymbol w}}:
&\|\hat{\boldsymbol{W}}^{(\mathrm{in}),N}_{k_l}\|_F\le\bar{\rho}_{\mathrm{inner}},\\
&\|\hat{\boldsymbol{W}}^{(\mathrm{in}),n}_{k_l,j_n}\|_F\le 1,\! \forall n<N,\\ &\|\hat{\boldsymbol{W}}^{(\mathrm{out}),L}\|_F\le\bar{\rho}_{\mathrm{outer}},
\ \|\hat{\boldsymbol{W}}^{(\mathrm{out}),l}_{j_l}\|_F\le 1
\Bigr\}.
\end{align*}
It therefore suffices to bound the empirical Rademacher complexity
$\mathcal{R}_{\mathcal{S}_x}(\hat{\mathcal{F}}_{G,\rho})$.
We consider the coordinate-wise complexity for the $M$-dimensional output and
analyze the scaled quantity $T\mathcal{R}_{\mathcal{S}_x}(\hat{\mathcal{F}}_{G,\rho})$
via a peeling argument adapted to the multi-block, multi-iteration structure,
\begin{align}
T\mathcal{R}
&:= T\mathcal{R}_{\mathcal{S}_x} (\hat{\mathcal{F}}_{G,\rho}) \nonumber \\
&\labeleq{a} \mathbb{E}_\xi \Bigg[
\sup_{\hat{\boldsymbol{w}}}
\sum_{k=1}^{K}\sum_{t=1}^T \sum_{m=1}^M
\xi_{k,t,m}\,
f_{\hat{\boldsymbol{w}}} (\boldsymbol{x}_{t})_{k,m}
\Bigg] \nonumber\\
&\labeleq{b} \mathbb{E}_\xi \Bigg[
\sup_{\hat{\boldsymbol{w}}}
\sum_{k=1}^{K}\sum_{m=1}^M
\sum_{t=1}^T \xi_{k,t,m}\nonumber\\
&\qquad\qquad\qquad\cdot\sum_{j_{N-1}}
\big\langle
\hat{\boldsymbol{W}}^{(\mathrm{in}),N}_{k_L,m,j_{N-1}},
\boldsymbol{z}^{(\mathrm{in}),N-1}_{k_L,j_{N-1}}(\boldsymbol{x}_{t})
\big\rangle
\Bigg] \nonumber\\
&\labeleq{c} \mathbb{E}_\xi \Bigg[
\sup_{\hat{\boldsymbol{w}}}
\sum_{k=1}^{K}\sum_{j_{N-1}}\sum_{m=1}^M
(\hat{\boldsymbol{W}}^{(\mathrm{in}),N}_{k_L,m,j_{N-1}})^\top\nonumber\\
&\qquad\qquad\qquad\cdot\sum_{t=1}^T
\xi_{k,t,m}
\boldsymbol{z}^{(\mathrm{in}),N-1}_{k_L,j_{N-1}}(\boldsymbol{x}_{t})
\Bigg] \nonumber\\
&\labelle{d}
\bar{\rho}_{\mathrm{inner}}
\mathbb{E}_\xi \Bigg[\!
\sum_{k=1}^{K}
\sup_{\hat{\boldsymbol{w}}}\!\!
\max_{m,j_{N-1}}
\Bigg\|
\sum_{t=1}^T
\xi_{k,t,m}
\boldsymbol{z}^{(\mathrm{in}),N-1}_{k_L,j_{N-1}}(\boldsymbol{x}_{t})
\Bigg\|_2
\Bigg] \nonumber\\
&\labelle{e}
\bar{\rho}_{\mathrm{inner}}
\mathbb{E}_\xi \Bigg[
K
\sup_{\hat{\boldsymbol{w}}}
\max_{\substack{k_L,m,\\j_{N-1}}}
\Bigg\|
\sum_{t=1}^T
\xi_{k_L,t,m}
\boldsymbol{z}^{(\mathrm{in}),N-1}_{k_L,j_{N-1}}(\boldsymbol{x}_{t})
\Bigg\|_2
\Bigg] \nonumber\\
&\labeleq{f}
\bar{\rho}_{\mathrm{inner}}K
\mathbb{E}_\xi \Bigg[
\sup_{\hat{\boldsymbol{w}}}
\max_{\substack{k_L,m,\\j_{N-1}}}
\Bigg\|
\sum_{t=1}^T
\xi_{k_L,t,m}
\boldsymbol{z}^{(\mathrm{in}),N-1}_{k_L,j_{N-1}}(\boldsymbol{x}_{t})
\Bigg\|_2
\Bigg].
\label{eq:11}
\end{align}
Here, $k_L$ denotes the terminal block index after unrolling the outer
iterations. Since each output coordinate corresponds to a unique terminal
block, we relabel the maximization index accordingly. Step $(d)$ follows from
the Frobenius-norm constraint on the output-layer weights of the inner MLP,
$\|\hat{\boldsymbol{W}}^{(\mathrm{in}),N}_{k_L}\|_F \le \bar{\rho}_{\mathrm{inner}}$.
By convexity, the supremum is attained by concentrating the norm on a single
row of $\hat{\boldsymbol{W}}^{(\mathrm{in}),N}_{k_L}$, and the bound follows from the
Cauchy--Schwarz inequality.

To bound the expectation, we apply the log-sum-exp inequality together with
Jensen's inequality: for any $\lambda>0$ and nonnegative random variable $Z$,
$\mathbb{E}[Z] \le \lambda^{-1}\log \mathbb{E}[\exp(\lambda Z)]$.
This yields
\begin{align}
&\mathbb{E}_\xi \bigg[
\sup_{\hat{\boldsymbol{w}}}
\max_{\substack{k_L,m,\\j_{N-1}}}
\bigg\|
\sum_{t=1}^T
\xi_{k_L,t,m}
\boldsymbol{z}^{(\mathrm{in}),N-1}_{k_L,j_{N-1}}(\boldsymbol{x}_{t})
\bigg\|_2
\bigg] \nonumber\\
&\le
\frac{1}{\lambda}
\log
\mathbb{E}_\xi
\sup_{\substack{\hat{\boldsymbol{w}},k_L,m,\\j_{N-1}}}
\exp\!\left(
\lambda
\left\|
\sum_{t=1}^T
\xi_{k_L,t,m}
\boldsymbol{z}^{(\mathrm{in}),N-1}_{k_L,j_{N-1}}(\boldsymbol{x}_{t})
\right\|_2
\right) \nonumber\\
&=
\frac{1}{\lambda}
\log
\mathbb{E}_\xi
\sup_{\substack{\hat{\boldsymbol{w}},k_L,m,\\j_{N-1}}}
\exp\!\Biggl(
\lambda
\sqrt{
\Bigg\|
\sum_{t=1}^T
\xi_{k_L,t,m}
\boldsymbol{z}^{(\mathrm{in}),N-1}_{k_L,j_{N-1}}(\boldsymbol{x}_{t})
\Bigg\|_2^2}
\Biggr).
\label{eq:12}
\end{align}
We then recursively apply the peeling argument of Lemma~\ref{lemma:2} with
$g(x)=\exp(\lambda x)$.
This is valid since the normalized weights satisfy
$\|\hat{\boldsymbol{W}}^{(\mathrm{in}),n}_{k_l,j_n}\|_F \le 1$ and the ReLU activation is positively
homogeneous and $1$-Lipschitz.
Starting from layer $N-1$, we obtain
\begin{align}
&\mathbb{E}_\xi
\sup_{\substack{\hat{\boldsymbol{w}},k_L,m,\\j_{N-1}}}
\exp\!\Biggl(
\lambda
\sqrt{
\bigg\|
\sum_{t=1}^T
\xi_{k_L,t,m}
\boldsymbol{z}^{(\mathrm{in}),N-1}_{k_L,j_{N-1}}(\boldsymbol{x}_{t})
\bigg\|_2^2}
\Biggr) \nonumber \\
&\labeleq{a}\!
\mathbb{E}_\xi\!\!\!\!
\sup_{\substack{\hat{\boldsymbol{w}},k_L,m,\\j_{N-1}}}
\!\!\!\!\exp\!\Biggl(
\!\!\lambda\!
\sqrt{
\!\!\bigg\|
\!\sum_{t=1}^T\!
\xi_{k_L,t,m}
\sigma\!\big(
\!\hat{\boldsymbol{W}}^{(\mathrm{in}),N-1}_{k_L,j_{N-1}}\!\!
\boldsymbol{v}^{(\mathrm{in}),N-2}_{k_L,j_{N-1}}\!\!(\boldsymbol{x}_{t})
\big)\!
\bigg\|_2^2}
\Biggr) \nonumber\\
&\labelle{b}
2\,
\mathbb{E}_\xi
\sup_{\substack{\hat{\boldsymbol{w}},k_L,m,\\j_{N-1}}}
\exp\!\Biggl(
\lambda
\sqrt{
\bigg\|
\sum_{t=1}^T
\xi_{k_L,t,m}
\boldsymbol{v}^{(\mathrm{in}),N-2}_{k_L,j_{N-1}}(\boldsymbol{x}_{t})
\bigg\|_2^2}
\Biggr) \nonumber\\
&\labeleq{c}
2\,
\mathbb{E}_\xi
\sup_{\substack{\hat{\boldsymbol{w}},k_L,m,\\j_{N-1}}}
\exp\!\Biggl(
\lambda
\Bigg[
\sum_{\substack{
j_{N-2}\in
\mathrm{pred}(N\!-\!1,j_{N-1})_{k_L}}}
\!\!\bigg\|
\sum_{t=1}^T
\xi_{k_L,t,m}\nonumber\\
&\qquad\qquad\qquad\cdot
\sigma\!\big(
\hat{\boldsymbol{W}}^{(\mathrm{in}),N-2}_{k_L,j_{N-2}}
\boldsymbol{v}^{(\mathrm{in}),N-3}_{k_L,j_{N-2}}(\boldsymbol{x}_{t})
\big)
\bigg\|_2^2
\Bigg]^{\frac12}
\Biggr) \nonumber\\
&\labelle{d}
4\,
\mathbb{E}_\xi
\sup_{\substack{\hat{\boldsymbol{w}},k_L,m,\\j_{N-1},j_{N-2}}}
\exp\!\Bigg(
\lambda
\Big[
|\mathrm{pred}(N\!-\!1,j_{N-1})_{k_L}|\nonumber\\
&\qquad\qquad\qquad\cdot
\bigg\|
\sum_{t=1}^T
\xi_{k,t,m}
\boldsymbol{v}^{(\mathrm{in}),N-3}_{k_L,j_{N-2}}(\boldsymbol{x}_{t})
\bigg\|_2^2
\Big]^{\frac12}
\Bigg)\nonumber\\
&\labelle{f}
2^N
\mathbb{E}_\xi
\max_{\substack{k_L,m, \\ j_0, \dots, j_N}}
\exp \Bigg(
\lambda
\Bigg[
\prod_{n=1}^{N-1}
\bigl|\mathrm{pred}(n, j_n)_{k_L}\bigr|\nonumber\\
&\qquad\qquad\qquad\cdot
\bigg\|
\sum_{t=1}^T
\xi_{k_L,t,m}
\boldsymbol{z}^{(\mathrm{in}),0}_{k_L,j_0}(\boldsymbol{x}_{t})
\bigg\|_2^2
\Bigg]^{\frac{1}{2}}
\Bigg)\nonumber\\
&\labelle{h} 2^N \mathbb{E}_\xi
\max_{\substack{\hat{\boldsymbol{w}},k_L,m, \\ j_0, \dots,j_N}}
\exp \Bigg(
\lambda
\Bigg[
\prod_{n=1}^{N-1}
\bigl|\mathrm{pred}(n, j_n)_{k_L}\bigr|\nonumber\\
&\qquad\qquad\cdot
\bigg\|
\sum_{t=1}^T
\xi_{k_L,t,m}
\sigma\big(
\hat{\boldsymbol{W}}^{(\mathrm{out}),L}_{k_L}
\boldsymbol{v}^{(\mathrm{out}),L-1}_{k_L}(\boldsymbol{x}_{t})
\big)
\bigg\|_2^2
\Bigg]^{\frac{1}{2}}
\Bigg) \nonumber\\
&\labelle{i}
2^{N+1}
\mathbb{E}_\xi
\max_{\substack{\hat{\boldsymbol{w}},m, \\ j_0, \dots,j_N, \\ k_L}}
\exp \Bigg(
\lambda \bar{\rho}_{\mathrm{outer}}
\sqrt{
\prod_{n=1}^{N-1}
\bigl|\mathrm{pred}(n, j_n)_{k_L}\bigr|}
\nonumber\\
&\qquad\qquad\qquad\cdot
\bigg\|
\sum_{t=1}^T
\xi_{k_L,t,m}
\boldsymbol{v}^{(\mathrm{out}),L-1}_{k_{L}}(\boldsymbol{x}_{t})
\bigg\|_2
\Bigg) \nonumber\\
&\labelle{j}
2^{N+1}
\mathbb{E}_\xi
\max_{\substack{\hat{\boldsymbol{w}},m, \\ j_0, \dots,j_N, \\ k_L,k_{L-1}}}
\exp \Bigg(
\lambda \bar{\rho}_{\mathrm{outer}}\nonumber\\
&\qquad\qquad\qquad\cdot\sqrt{
\prod_{n=1}^{N-1}
\bigl|\mathrm{pred}(n, j_n)_{k_L}\bigr|
\cdot
\bigl|\mathrm{pred}(L,k_L)\bigr|}
\nonumber\\
&\qquad\qquad\qquad\cdot
\bigg\|
\sum_{t=1}^T
\xi_{k_{L},t,m}
\boldsymbol{z}^{(\mathrm{out}),L-1}_{k_{L-1}}(\boldsymbol{x}_{t})
\bigg\|_2
\Bigg)\nonumber\\
&\labelle{k}
2^{(N+1)L}
\mathbb{E}_\xi
\max_{\substack{m, j_0,\dots,j_N \\ k_0,\dots,k_L}}
\exp \Bigg(
\lambda \bar{\rho}_{\mathrm{outer}}
(\bar{\rho}_{\mathrm{inner}})^{L-1}\nonumber\\
&\qquad\qquad\qquad\cdot
\sqrt{
\prod_{l=1}^{L}
\bigl|\mathrm{pred}(l,k_l)\bigr|
\cdot
\prod_{l=1}^{L}
\prod_{n=1}^{N-1}
\bigl|\mathrm{pred}(n,j_n)_{k_l}\bigr|}
\nonumber\\
&\qquad\qquad\qquad\qquad\cdot
\bigg\|
\sum_{t=1}^T
\xi_{k_L,t,m}
\boldsymbol{z}^{(\mathrm{out}),0}_{k_0}(\boldsymbol{x}_{t})
\bigg\|_2
\Bigg).
\label{eq:15}
\end{align}
Here, $(f)$ follows from recursively applying the peeling argument over the
$N$-layer inner MLPs, while $(h)$--$(j)$ result from applying
Peeling Lemma~\ref{lemma:2} to the outer network. Step $(k)$ is obtained by
iterating the outer peeling process over $L$ blocks.
The maximization is taken over all valid unit-level paths
$(j_0,\dots,j_N)$ and block-level paths $(k_0,\dots,k_L)$ satisfying
$j_{n-1}\in\mathrm{pred}(n,j_n)_{k_l}$ and
$k_{l-1}\in\mathrm{pred}(l,k_l)$.
Since all index sets are finite, the supremum reduces to a maximum.
Each block outputs an $M$-dimensional vector, yielding a factor $M^L$ via a union
bound over output coordinates. We notice that:
\begin{align}
    & \mathbb{E}_\xi \max_{ \substack{m,j_0, \dots,j_N}}\;\max_{k_{0},\dots,k_{L}}\exp \Bigg( \lambda  \bar{\rho}_{\mathrm{outer}} (\bar{\rho}_{\mathrm{inner}})^{L-1} \nonumber\\
    &\qquad \cdot\; \sqrt{\prod_{l=1}^{L} |\mathrm{pred}(l, k_l)|\cdot\prod_{l=1}^{L}\left( \prod_{n=1}^{N-1} |\mathrm{pred}(n, j_n)_{k_l}| \right)}\nonumber\\
    &\qquad\cdot\; \bigg\| \sum_{t=1}^T \xi_{k_L,t,m} 
    \boldsymbol{z}^{(\mathrm{out}),0}_{k_{0}}(\boldsymbol{x}_{t})
    \bigg\|_2 \Bigg)\nonumber\\
    & \leq \sum_{m,j_0, \dots,j_N} \sum_{k_{0},\dots,k_{L}}\mathbb{E}_\xi \exp \Bigg(\lambda \bar{\rho}_{\mathrm{outer}} (\bar{\rho}_{\mathrm{inner}})^{L-1}\nonumber\\
    &\qquad \cdot\; \sqrt{\prod_{l=1}^{L} |\mathrm{pred}(l, k_l)|\cdot\prod_{l=1}^{L}\left( \prod_{n=1}^{N-1} |\mathrm{pred}(n, j_n)_{k_l}| \right)}\nonumber\\
    &\qquad\cdot\; \bigg\| \sum_{t=1}^T \xi_{k_L,t,m} 
    \boldsymbol{z}^{(\mathrm{out}),0}_{k_{0}}(\boldsymbol{x}_{t})
    \bigg\|_2 \Bigg)\nonumber\\
    & \leq M^L \prod_{l=1}^{L}\left( \prod_{n=1}^{N-1}  d^{(\mathrm{in})}_{l,n} \right) \cdot\prod_{l=1}^{L}  d^{(\mathrm{out})}_{{l}}\max_{ \substack{m,j_0, \dots,j_N}}\;\max_{k_{0},\dots,k_{L}}\mathbb{E}_\xi \nonumber\\
    &\qquad\exp \Bigg( \lambda \bar{\rho}_{\mathrm{outer}} (\bar{\rho}_{\mathrm{inner}})^{L-1}\nonumber\\
    &\qquad \cdot\; \sqrt{\prod_{l=1}^{L} |\mathrm{pred}(l, k_l)|\cdot\prod_{l=1}^{L}\left( \prod_{n=1}^{N-1} |\mathrm{pred}(n, j_n)_{k_l}| \right)}\nonumber\\
    &\qquad\cdot\; \bigg\| \sum_{t=1}^T \xi_{k_L,t,m} 
    \boldsymbol{z}^{(\mathrm{out}),0}_{k_{0}}(\boldsymbol{x}_{t})
    \bigg\|_2 \Bigg).
\label{eq:17}
\end{align}
By combining the peeling bounds with exponential-moment arguments, the empirical
Rademacher complexity admits:
\begin{align}
    &T\mathcal{R}_{\mathcal{S}_x} (\hat{\mathcal{F}}_{G,\bar{\rho}}) \nonumber\\
   &\leq \bar{\rho}_{\mathrm{inner}} K \mathbb{E}_\xi \bigg[ \sup_{\hat{\boldsymbol{w}}} \max_{\substack{k_L,m,\\j_{N-1}}} \bigg\| \sum_{t=1}^T \xi_{k_L,t,m}\boldsymbol{z}^{(\mathrm{in}),N-1}_{k_L,j_{N-1}}(\boldsymbol{x}_{t}) \bigg\|_2 \bigg]\nonumber\\
    &\leq \frac{\bar{\rho}_{\mathrm{inner}}\; K}{\lambda} \log \Bigg(
    2^{(N+1)L}\, M^L
    \prod_{l=1}^{L}\Big( \prod_{n=1}^{N-1}  d^{(\mathrm{in})}_{l,n} \Big)
    \prod_{l=1}^{L}  d^{(\mathrm{out})}_{l} \nonumber\\
    &\qquad\cdot\max_{ \substack{m,j_0, \dots,j_N}}\;\max_{k_{0},\dots,k_{L}}\
    \mathbb{E}_\xi \exp \Bigg(
    \lambda\, \bar{\rho}_{\mathrm{outer}}
    (\bar{\rho}_{\mathrm{inner}})^{L-1}\nonumber\\
    &\qquad\qquad\cdot
    \sqrt{
    \prod_{l=1}^{L} |\mathrm{pred}(l, k_l)|
    \prod_{l=1}^{L}\Big( \prod_{n=1}^{N-1}
    |\mathrm{pred}(n, j_n)_{k_l}| \Big)
    } \nonumber\\
    &\qquad\qquad\cdot
    \bigg\|
    \sum_{t=1}^T
    \xi_{k_L,t,m}\,
    \boldsymbol{z}^{(\mathrm{out}),0}_{k_{0}}(\boldsymbol{x}_{t})
    \bigg\|_2
    \Bigg)
    \Bigg).
\end{align}
The remaining expectation is of the form
$\mathbb{E}_\xi \exp\!\big(\alpha\|\sum_{t=1}^T\xi_t\boldsymbol{z}_t\|_2\big)$.
By standard sub-Gaussian concentration
(e.g.,~\cite[Thm.~6.2]{Boucheron2013Concentration}),
for any $\alpha>0$,
\begin{align}
    &\mathbb{E}_\xi \exp \Bigg(
    \alpha \bigg\|
    \sum_{t=1}^T \xi_{k_L,t,m}\,
    \boldsymbol{z}^0_{j_0}(\boldsymbol{x}_t)
    \bigg\|_2
    \Bigg) \nonumber\\
    &\leq
    \exp \!\!\Bigg(
    \frac{\alpha^2}{2}
    \sum_{t=1}^T
    \big\|
    \boldsymbol{z}^{(\mathrm{out}),0}_{k_{0}}(\boldsymbol{x}_{t})
    \big\|_2^2
    + \alpha
    \sqrt{
    \sum_{t=1}^T
    \big\|
    \boldsymbol{z}^{(\mathrm{out}),0}_{k_{0}}(\boldsymbol{x}_{t})
    \big\|_2^2
    }
    \Bigg).
\label{eq:18}
\end{align}
By combining \eqref{eq:11}--\eqref{eq:18} and setting
\begin{align*}
\alpha
\!=\!
\lambda \bar{\rho}_{\mathrm{outer}}
(\bar{\rho}_{\mathrm{inner}})^{L-1}\!\!\!
\!\sqrt{\!
\prod_{l=1}^{L} \!|\mathrm{pred}(l,k_l)|\!
\prod_{l=1}^{L}\!\prod_{n=1}^{N-1}\!
|\mathrm{pred}(n,j_n)_{k_l}|
},
\end{align*}
we obtain the following bound:
\begin{align}
&T \mathcal{R}_{\mathcal{S}_x}(\hat{\mathcal{F}}_{G,\bar{\rho}})\nonumber\\
&\leq
\frac{\bar{\rho}_{\mathrm{inner}} K}{\lambda}
\Bigg[
(N+1)L\log 2
+ L\log M
\nonumber\\
&\qquad+ \sum_{l=1}^{L}\log d^{(\mathrm{out})}_l+ \sum_{l=1}^{L}\sum_{n=1}^{N-1}\log d^{(\mathrm{in})}_{l,n}
\Bigg] \nonumber\\
&\quad+ \frac{\lambda K}{2}
\bar{\rho}_{\mathrm{outer}}^{2}
(\bar{\rho}_{\mathrm{inner}})^{2(L-1)}\!\!\max_{k_{0},\dots,k_L}
\prod_{l=1}^{L} |\mathrm{pred}(l,k_l)|
 \nonumber\\
&\qquad\cdot\!\!\max_{j_0,\dots,j_N}
\prod_{l=1}^{L}\prod_{n=1}^{N-1}
|\mathrm{pred}(n,j_n)_{k_l}|
\max_{k_{0}}
\sum_{t=1}^{T}
\big\|
\boldsymbol{z}^{(\mathrm{out}),0}_{k_{0}}(\boldsymbol{x}_t)
\big\|_2^2 \nonumber\\
&\quad+ K\bar{\rho}_{\mathrm{outer}}
(\bar{\rho}_{\mathrm{inner}})^{L}
\Bigg[
\max_{k_{0},\dots,k_L}
\prod_{l=1}^{L} |\mathrm{pred}(l,k_l)|
\nonumber\\
&\qquad\!\!\!\cdot\!\!\!\max_{j_0,\dots,j_N}\!
\prod_{l=1}^{L}\!\prod_{n=1}^{N-1}
\!|\mathrm{pred}(n,j_n)_{k_l}| 
\max_{k_{0}}
\sum_{t=1}^{T}
\big\|
\boldsymbol{z}^{(\mathrm{out}),0}_{k_{0}}(\boldsymbol{x}_t)
\big\|_2^2\!
\Bigg]^{\!\frac12}.
\label{eq:19}
\end{align}
Optimizing the bound with respect to $\lambda$ yields
\begin{align*}
    \lambda &= \Bigg[ 2 \Bigg((N+1)L\log 2
+ L\log M+ \sum_{l=1}^{L}\log d^{(\mathrm{out})}_l
\\
&\qquad+ \sum_{l=1}^{L}\sum_{n=1}^{N-1}\log d^{(\mathrm{in})}_{l,n} \Bigg) \bigg/ \Bigg(\bar{\rho}_{\mathrm{inner}}^{\;2(L-1)}\bar{\rho}^{2}_{\mathrm{outer}} \\
    &\qquad\cdot\!\!\max_{k_{0},\dots,k_L}
\prod_{l=1}^{L} |\mathrm{pred}(l,k_l)|
\max_{j_0,\dots,j_N}\!
\prod_{l=1}^{L}\!\prod_{n=1}^{N-1}
\!|\mathrm{pred}(n,j_n)_{k_l}| \\
&\qquad\cdot
\max_{k_{0}}
\sum_{t=1}^{T}
\big\|
\boldsymbol{z}^{(\mathrm{out}),0}_{k_{0}}(\boldsymbol{x}_t)
\big\|_2^2 \Bigg) \Bigg]^{\frac{1}{2}}.
\end{align*}
Substituting this choice of $\lambda$ into~\eqref{eq:19} completes the proof.
\hfill\IEEEQED
\subsection{Proof of Theorem~\ref{thm:1}}
\label{app:The 4.5}
For \(r \in \mathbb{N}\cup\{0\}\), define function class $\mathcal{F}_{G,r}$.
We assign a failure probability budget
\begin{align*}
\delta_r \coloneqq \frac{\delta}{r(r+1)}, \qquad r \ge 1,
\end{align*}
which satisfies \(\sum_{r\ge1} \delta_r = \delta\).

By Lemma~\ref{lemma:1}, for any fixed shell index \(r\), the following inequality holds
with probability at least \(1-\delta_r\):
\begin{align}
\label{eq:per-r}
\text{err}_P(\boldsymbol{w}) - \text{err}_{\mathcal{S}}^\gamma(\boldsymbol{w})
\;\le\;
\frac{2\sqrt{2}}{\gamma}\, \mathcal{R}_{\mathcal{S}_x}(\mathcal{F}_{G,r})
\;+\;
3\sqrt{\frac{\log\!\bigl(2/\delta_r\bigr)}{2T}}.
\end{align}
Substituting the Rademacher complexity upper bound from
Proposition~\ref{prop:1}, we introduce the shorthand notation
\begin{align*}
A &\coloneqq (N+1)L\log 2
+ L\log M
\\
&\qquad+ \sum_{l=1}^{L}\log d^{(\mathrm{out})}_l
+ \sum_{l=1}^{L}\sum_{n=1}^{N-1}\log d^{(\mathrm{in})}_{l,n},\\
B &\coloneqq
\max_{k_{0},\dots,k_L}
\prod_{l=1}^{L} |\mathrm{pred}(l,k_l)|
\max_{j_0,\dots,j_N}
\prod_{l=1}^{L}\prod_{n=1}^{N-1}
|\mathrm{pred}(n,j_n)_{k_l}| \\
&\qquad\cdot
\max_{k_{0}}
\sum_{t=1}^{T}
\big\|
\boldsymbol{z}^{(\mathrm{out}),0}_{k_{0}}(\boldsymbol{x}_t)
\big\|_2^2.
\end{align*}
Proposition~\ref{prop:1}, together with the uniform inner-block bound
\(\bar\rho_{\mathrm{inner}}\) and outer-block bound
\(\bar\rho_{\mathrm{outer}}\), yields
\begin{align*}
\mathcal{R}_{\mathcal{S}_x}(\mathcal{F}_{G,r})
\;\le\;
\frac{\bar\rho_{\mathrm{inner}}^{\,L}\bar\rho_{\mathrm{outer}}\,K}{T}
\,(1+\sqrt{2A})\,\sqrt{B}.
\end{align*}

Plugging this bound into~\eqref{eq:per-r} implies that, with probability at least
\(1-\delta_r\),
\begin{align*}
\text{err}_P(\boldsymbol{w})-\text{err}_{\mathcal{S}}^\gamma(\boldsymbol{w})
\;\le\;
&\frac{2\sqrt{2}\,K\,\bar\rho_{\mathrm{inner}}^{\,L}\bar\rho_{\mathrm{outer}}}
{\gamma T}
(1+\sqrt{2A})\sqrt{B}
\\
&\quad
+3\sqrt{\frac{\log\!\bigl(2/\delta_r\bigr)}{2T}}.
\end{align*}
Applying the union bound over all \(r \ge 1\) (since
\(\sum_{r}\delta_r = \delta\)), the above inequality holds uniformly for all
parameter settings with probability at least \(1-\delta\).
For any \(f_{\boldsymbol{w}}\in\mathcal{F}_{G,r}\) with norm \(\rho(\boldsymbol{w})\),
which is upper-bounded by
\(\bar\rho_{\mathrm{inner}}^{\,L}\bar\rho_{\mathrm{outer}}\),
we choose \(r = \lceil \rho(\boldsymbol{w}) \rceil\ge 1\).
Noting that
\begin{align*}
\log\!\Big(\frac{2}{\delta_r}\Big)
= \log\!\Big(\frac{2r(r+1)}{\delta}\Big)
\le \log\!\Big(\frac{2(r+1)^2}{\delta}\Big),
r \!\le\! \!\rho(\boldsymbol{w}) \!+\! 1,
\end{align*}
we obtain that, with probability at least \(1-\delta\),
\begin{align}
\text{err}_P(\boldsymbol{w}) - \text{err}_{\mathcal{S}}^\gamma(\boldsymbol{w})
&\le
\frac{2\sqrt{2}\,K\,\lceil\rho(\boldsymbol{w})\rceil}{\gamma T}
(1+\sqrt{2A})\sqrt{B} \notag\\
&\quad
+ 3\sqrt{\frac{
\log\!\big(2\,\lceil\rho(\boldsymbol{w})\rceil(\lceil\rho(\boldsymbol{w})\rceil+1)/\delta\big)
}{2T}} \notag\\
&\le
\frac{2\sqrt{2}\,K\,(\rho(\boldsymbol{w})+1)}{\gamma T}
(1+\sqrt{2A})\sqrt{B} \notag\\
&\quad
+ 3\sqrt{\frac{\log\!\big(2(\rho(\boldsymbol{w})+2)^2/\delta\big)}{2T}},
\label{eq:theorem45-final}
\end{align}
which completes the proof. \hfill \IEEEQED
\bibliographystyle{IEEEtran}
\bibliography{paper}
\end{document}